\documentclass{aastex631}

\usepackage{acronym}
\usepackage{xcolor}
\usepackage{graphicx}
\usepackage{savesym}
\savesymbol{tablenum}
\usepackage{siunitx}
\usepackage{units}
\restoresymbol{SIX}{tablenum}
\usepackage[T1]{fontenc}
\usepackage[colorlinks=true]{hyperref}
\usepackage{makecell}
\graphicspath{{./}{figures/}}

\DeclareSIUnit[number-unit-product = {}] \erg{erg}
\DeclareSIUnit[number-unit-product = {}] \kpc{kpc}

\newcommand{\hz}{\, \si{\hertz}}

\newcommand*{\euler}{\mathrm{e}}

\newcommand*{\equationmention}[1]{equation~(#1)}
\newcommand*{\eqref}[1]{equation~(\ref{#1})}
\newcommand*{\tableref}[1]{Table~\ref{#1}}
\newcommand*{\figureref}[1]{Figure~\ref{#1}}
\newcommand*{\subsectionref}[1]{Sec.~\ref{#1}}
\newcommand*{\appendixref}[1]{Appendix~\ref{#1}}
\newcommand*{\Alfven}{Alfv\'{e}n}

\acrodef{aLIGO}{Advanced LIGO}
\acrodef{FBS}{fraction of background segments}
\acrodef{fmode}[$f$-mode]{fundamental oscillation mode}
\acrodef{gmode}[$g$-mode]{buoyancy mode}
\acrodef{GW}[gravitational wave]{gravitational wave}
\acrodef{hrss}[$h_\mathrm{rss}$]{root-sum-square of the integrated gravitational-wave strain}
\acrodef{EGW}[$E_\mathrm{GW}$]{gravitational-wave energy}
\acrodef{LHO}{LIGO Hanford Observatory}
\acrodef{LLO}{LIGO Livingston Observatory}
\acrodef{O2}{second observing run}
\acrodef{O3}{third observing run}
\acrodef{QPO}{quasi-periodic oscillation}
\acrodef{S6}{LIGO's sixth science run}
\acrodef{SNR}{Signal to noise ratio}
\acrodef{STAMP}[STAMP]{Stochastic Transient Analysis Multi-detector Pipeline}
\acrodef{tf}[time-frequency]{time-frequency}

\newcommand{\definiteIntegral}[4]{
	\int_{#1}^{#2} \! #3 \, \mathrm{d}#4
}

\newcommand{\hrssulF}{$h_\mathrm{rss}^{50\%}$}
\newcommand{\egwulFplain}{E_\mathrm{GW}^{50\%}}
\newcommand{\egwulF}{$\egwulFplain$}
\newcommand*{\hrss}{h_\mathrm{rss}}
\newcommand*{\EEMISO}{E_\mathrm{EM}^\mathrm{iso}}
\newcommand*{\Eratio}{\frac{\egwulFplain}{\EEMISO}}
\newcommand*{\Eratioflat}{\egwulFplain/\EEMISO}

\begin{document}

\let\myauthor\author
\let\myaffiliation\affiliation
\let\mymaketitle\maketitle

\title{Search for gravitational-wave transients associated with magnetar bursts in Advanced LIGO and Advanced Virgo data from the third observing run}
\author{The LIGO Scientific Collaboration, the Virgo Collaboration, and the KAGRA Collaboration}

\begin{abstract}
  Gravitational waves are expected to be produced from neutron star oscillations associated with magnetar giant flares and short bursts. We present the results of a search for short-duration (milliseconds to seconds) and long-duration ($\sim$ \unit[100]{s}) transient gravitational waves from 13 magnetar short bursts observed during Advanced LIGO, Advanced Virgo and KAGRA's third observation run. 
  These 13 bursts come from two magnetars, SGR 1935$+$2154 and Swift J1818.0$-$1607.
  We also include three other electromagnetic burst events detected by Fermi GBM which were identified as likely coming from one or more magnetars, but they have no association with a known magnetar. No magnetar giant flares were detected during the analysis period.
  We find no evidence of gravitational waves associated with any of these 16 bursts.
  We place upper bounds on the root-sum-square of the integrated gravitational-wave strain that reach \unit[$2.2 \times 10^{-23}$]{$/\sqrt{\text{Hz}}$} at \unit[100]{Hz} for the short-duration search and \unit[$8.7 \times 10^{-23}$]{$/\sqrt{\text{Hz}}$} at \unit[450]{Hz} for the long-duration search, given a detection efficiency of 50\%.
  For a ringdown signal at \unit[1590]{Hz} targeted by the short-duration search the limit is set to \unit[$1.8 \times 10^{-22}$]{$/\sqrt{\text{Hz}}$}.
  Using the estimated distance to each magnetar, we derive upper bounds on the emitted gravitational-wave energy of \unit[$3.2 \times 10 ^{43}$]{erg} (\unit[$7.3 \times 10^{43}$]{erg}) for SGR 1935$+$2154 and \unit[$8.2 \times 10^{42}$]{erg} (\unit[$2.8 \times 10^{43}$]{erg}) for Swift J1818.0$-$1607, for the short-duration (long-duration) search.
  Assuming isotropic emission of electromagnetic radiation of the burst fluences, we constrain the ratio of gravitational-wave energy to electromagnetic energy for bursts from SGR 1935$+$2154 with available fluence information.
  The lowest of these ratios is $3 \times 10 ^3$.
\end{abstract}

\acresetall
\acused{GW}
\section{Introduction}

Magnetars---highly-magnetized neutron stars---exhibit intermittent bursts of hard X-rays and soft gamma rays with typical peak luminosities~$\lesssim10^{43}\,{\rm erg\,s}^{-1}$~\cite[see][for a review]{Kaspi_2017}.
Galactic \citep{evans80,hurley99,hurley05,mereghetti05,boggs07} and extragalactic \citep{mazets08,Abadie_2012,svinkin21,Burns_2021} giant flares have peak luminosities up to five orders of magnitude larger.
Although the mechanisms that causes these bursts and giant flares are not well understood, many models predict accompanying gravitational-wave emission from excited core and/or crust oscillations~\citep{Ioka2001,corsi_owen_2011,kashiyama11,zink12,ciolfi11}. This hypothesis is enhanced by the identification of \acp{QPO} in the tails of giant flares~\citep{barat83,israel05,strohmayer05,strohmayer06,watts06}, and possibly fainter bursts~\citep{huppenkothen14a, huppenkothen14b}, which have been attributed to various oscillations of the stellar crust and core~\cite[e.g.,][]{duncan98,messios01,piro05,strohmayer06,glampedakis06,levin07,colaiuda11, glampedakis14}.

Initial estimates for the potential gravitational-wave emission from giant flares were optimistic that the \ac{fmode} of magnetars could be excited by the catastrophic rearrangement of the star's interior magnetic field.
These \acp{fmode} are excited at frequencies between $\sim1\,{\rm kHz}$ and $3\,{\rm kHz}$, and could potentially emit up to $\sim10^{49}\,{\rm erg}$~\citep{Ioka2001,corsi_owen_2011} for $\sim100\,{\rm ms}$~\citep{lindblom83,mcdermott88,Wen_2019}, making them detectable with Advanced LIGO~\citep{LIGO}, Advanced Virgo~\citep{Virgo}, and KAGRA~\citep{KAGRA, KAGRA_design_overview_2021} gravitational-wave observatories~\citep{LVK_prospects2018}.
Detailed calculations of the rearrangement of the neutron star's magnetic field using analytic calculations~\citep{Levin2011} and numerical-relativity simulations~\citep{ciolfi11,zink12,ciolfi12,tsokaros21} yield more realistic estimates for the gravitational-wave energy emitted in the \ac{fmode} during these events. These models suggest \ac{GW}s associated with Galactic magnetar flares are not observable with the current generation of observatories, but instead require the sensitivity of at least third-generation observatories such as the Einstein Telescope~\citep{EinsteinTelescope} or Cosmic Explorer~\citep{reitze2019cosmic,Evans:2021gyd}, or dedicated kilohertz facilities such as the proposed NEMO observatory~\citep{NEMO}.

Other oscillation modes in the star, which are generally longer-lived and at lower frequencies than the \ac{fmode}, may be excited.
These include \acp{gmode} and \Alfven{} modes where the magnetic field provides the oscillation restoring force.
The latter of these modes has been suggested as potential long-lived gravitational-wave sources from magnetar giant flares \citep{kashiyama11,zink12}, with the resonant frequency correlated to the initial rise time of the magnetar giant flare \citep{mazets08,hurley99,hurley05}. 
The damping time and mode excitation amplitude of these modes are largely unknown, making them interesting candidates for longer-lived gravitational-wave signals.

Only one Galactic giant flare has been observed coincident with a LIGO-Virgo observing run: the 2004 giant flare from SGR 1806$-$20.
Upper limits on the gravitational-wave energy associated with the burst were determined to be between $\sim10^{46}$ and $10^{52}\,{\rm erg}$, depending on the assumed waveform model~\citep{Kalmus_2007,PhysRevLett.101.211102,kalmus2009search}.
A stacked search was performed on a series of bursts from the same magnetar occurring in the same minute of time, reducing the above upper limits to $\sim10^{45}$ and $10^{50}\,{\rm erg}$ assuming the same \ac{fmode} frequency and damping time for each burst \citep{PhysRevD.80.042001,Abbott_2009,kalmus2009search}.
Further searches for \acp{fmode} on 1279 bursts from six different magnetars yielded upper limits between $\sim10^{44}$ and $10^{47}\,{\rm erg}$~\citep{Abadie_2011}, while the first search for \acp{fmode} in the Advanced era of gravitational-wave interferometers yielded comparable upper limits on four bursts from two magnetars~\citep{Abbott_2019,schale2019search}.
These energy upper limits are in the range of possible gravitational-wave energies given the most optimistic predictions~\citep{Ioka2001,corsi_owen_2011}.

Longer-duration searches initially targeted the \ac{QPO} frequencies in the tail of the giant flare of SGR 1806$-$20, with upper limits of various modes at $\lesssim10^{46}\,{\rm erg}$~\citep{Matone_2007, PhysRevD.76.062003}, comparable to the electromagnetic energy emitted from the giant flare.
A study was performed on a method to detect \acp{GW} targeting repeated \acp{QPO}~\citep{PhysRevD.87.103008}.
A search for long-duration \acp{GW} from four magnetar bursts was performed using \ac{S6} data \citep{quitzowjames2016thesis, Quitzow_James_2017}.
The best upper limits from magnetar bursts come from recent observations in the Advanced LIGO-Virgo \ac{O2}, with gravitational-wave energies constrained to less than $\sim10^{44}$ to $10^{48}\,{\rm erg}$, again depending on signal model, frequency, and damping time~\citep{Abbott_2019,schale2019search}.
We previously placed limits on gravitational-wave emission from purported extragalactic magnetar giant flares~\citep{Abbott_2008, Abadie_2012}\footnote{GRB 070222, included in a search targeting gamma ray bursts~\citep{PhysRevLett.113.011102}, was later determined to likely be an extragalactic magnetar giant flare~\citep{Burns_2021,macquet21}.}, with long-duration searches constraining the gravitational-wave energy emitted to between $\sim10^{49}$ and $10^{52}\,{\rm erg}$ for four different giant flares~\citep{macquet21}.

In this Paper, we report on a search for \acp{GW} coincident with 13 magnetar short bursts from SGR 1935$+$2154, Swift J1818.0$-$1607, and 3 electromagnetic bursts from an unidentified source (or sources) during the LIGO-Virgo-KAGRA \ac{O3}.
Targeted gravitational-wave searches associated with magnetar bursts can broadly be split in two categories following theoretical predictions of short-duration and long-duration signals.
We describe a short-duration and long-duration search and provide upper limits on gravitational-wave strain for each search.  For the remainder of this paper, the term `upper limit' should be interpreted as meaning the threshold of a parameter at which 50\% of the signals are detected.

\section{Methodology}\label{sec:methodology}

The \ac{O3} observing run extended from April 1, 2019, to March 27, 2020,
with 3 gravitational-wave detectors taking data: \ac{LHO}, \ac{LLO} and Virgo, all of which had been upgraded so that the binary neutron star inspiral ranges increased by a factor of $1.53$ for \ac{LLO}, $1.64$ for \ac{LHO} and $1.73$ for Virgo \citep{Davis_2021} compared to their performance during \ac{O2}.
For each detector, several data quality checks are performed to mitigate terrestrial noise \citep{Davis_2021}.
In addition, multi-detector analyses are used to mitigate non-astrophysical features present in the data.
In January 2020, a new technique was implemented to mitigate the impact in LIGO detectors of scattered light, a transient noise coupled with ground motion.
Data available for bursts that occurred in February and March show a lower transient noise rate. 

The list of magnetar short bursts and giant flares from \cite{IPN_masterlist} provides the source object and observation time for each burst.
In both the long-duration and short-duration searches, we describe the data in which we look for a signal as the \textit{on-source} window, while the time around this composing the background is the \textit{off-source} window.
Analysis requirements include at least two gravitational-wave detectors in observation mode,
sufficiently good data quality,
and sufficient data available in the burst's on-source window.
Several bursts did not meet the two-detector criterion
(one detector available for 2651, 2658, 2659, and 2667 in \cite{IPN_masterlist} and
none for 2662, 2663, and 2664),
had poor data quality (2650 and 2672),
or had very little data available in the on-source window (2650).
Considering these requirements, 12 magnetar short bursts from known sources and three electromagnetic bursts thought to likely be magnetar short bursts from an unknown source or sources occurred when at least two detectors were in observation mode with sufficiently good data quality and are included in this search.
In addition, although burst 2651 occurred when \ac{LHO} was in observation mode and 87 \si{\s} before Virgo was in observation mode (\ac{LLO} was not taking data),
data of sufficiently good quality was available for most of the long-duration search on-source window, thus the burst was 
analyzed by the long-duration search only.
Eleven short bursts were from SGR 1935$+$2154 \citep{cummings2014newly}, a magnetar which emitted a fast radio burst in April of 2020 \citep{Kirsten_2020}, and two short bursts were from Swift J1818.0$-$1607, a magnetar discovered in 2020 \citep{evans2020swift}.
Note that SGR 1935$+$2154 generated more bursts on November 4, 2019, but only nine can be analyzed as at least two detectors in observing mode are necessary.

The source or sources of the remaining three electromagnetic bursts, detected by Fermi GBM \citep{meegan2009}, are unknown as they have very poor sky localization.
These three electromagnetic bursts were accompanied by a fourth burst which did not meet the two-detector condition necessary for our analysis, but all four of these bursts occurred in a 33 hour window of time between February 3 and February 4, 2020. 
Because of their temporal proximity, we search for a signal assuming that they were emitted by the same magnetar.
Only two known galactic magnetars are in the $3\sigma$ error region of all four electromagnetic bursts.
To obtain the best constraints,
we consider the closest of these two magnetars, 1 RXS J170849 at $3.8$ \si{\kpc} \citep{Durant2006}, as the source of these bursts.
The sources, times, active detectors, and the isotropic electromagnetic energy ($\EEMISO$) of each burst included in this search are listed in \tableref{tab:table_triggers};
fluence information for some of the bursts from SGR 1935$+$2154 were obtained from \cite{lin2020fermi} and used to estimate $\EEMISO$.

\renewcommand{\arraystretch}{1}

\begin{table*}
\begin{center}
\begin{tabular}{| c c c c c c c | }
  \hline
    Burst & Source & Date &  Time   (UTC) & Detectors & $\EEMISO$ (\si{\erg}) & GCN Circulars \\[0.5ex]
  \hline\hline
    2651 & SGR 1935$+$2154 & Nov 04, 2019 & 01:54:37 & H V* & - & 26169\\
  \hline
    2652 & SGR 1935$+$2154 & Nov 04, 2019 & 02:53:31 & H V & $1.4 \times 10^{39}$ & 26163, 26151\\
  \hline
    2653 & SGR 1935$+$2154 & Nov 04, 2019 & 04:26:55 & H L V & $1.1 \times 10^{39}$ & 26163\\
  \hline
    2654 & SGR 1935$+$2154 & Nov 04, 2019 & 06:34:00 & H L V & - & 26153\\
  \hline
    2655 & SGR 1935$+$2154 & Nov 04, 2019 & 09:17:53 & H L & $5.7 \times 10^{39}$  & 26163, 26154\\
  \hline
    2656 & SGR 1935$+$2154 & Nov 04, 2019 & 10:44:26 & H L & $ 2.2 \times 10^{40}$ & 26242, 26163, 26158, 26157\\
  \hline
    2657 & SGR 1935$+$2154 & Nov 04, 2019 & 12:38:38 & H L V & $2.7 \times 10^{39} $ & 26163 \\
  \hline
    2660 & SGR 1935$+$2154 & Nov 04, 2019 & 15:36:47 & H V & $ 1.2 \times 10^{39} $ & - \\
  \hline
    2661 & SGR 1935$+$2154 & Nov 04, 2019 & 20:29:39 & H V & $1.3 \times 10^{39}$ & 26165, 26166\\
  \hline
    2665 & SGR 1935$+$2154 & Nov 05, 2019 & 06:11:08 & H V & $7.8 \times 10^{40}$ & 26242\\
  \hline
    2668 & SGR 1935$+$2154 & Nov 15, 2019 & 20:48:41 & L V & $7.7 \times 10^{38}$ & - \\
  \hline
    2669 & - & Feb 03, 2020 & 03:17:11 & H L V & - & 26980\\
  \hline
    2670 & - & Feb 03, 2020 & 03:44:03 & H L V & - & 26969, 26980\\
  \hline
    2671 & - & Feb 03, 2020 & 20:39:37 & H L V & - & 26980\\
  \hline
    2673 & Swift J1818.0$-$1607 & Feb 28, 2020 & 22:19:32 & L V & - & - \\
  \hline
    2674 & Swift J1818.0$-$1607 & Mar 12, 2020 & 21:16:47 & H L V & - & 27373\\
  \hline
\end{tabular}
        \caption{\label{tab:table_triggers} List of magnetar short bursts considered for this search from the interplanetary network master burst list \citep{IPN_masterlist}, and the available GW detectors: LIGO Hanford Observatory (H), LIGO Livingston Observatory (L), Virgo (V). Three electromagnetic bursts (2669, 2670, 2671) were not identified with a source and are thought to likely come from one or more magnetar(s), as reported in GCN circular 26980 \citep{GCN26980} and in the Fermi GBM \cite{meegan2009} on-board Trigger Catalog,  available online$^{\dagger}$.
The isotropic electromagnetic energy ($\EEMISO$) is estimated from the fluence given in \cite{lin2020fermi} considering a distance of \unit[9.0]{kpc} for SGR 1935$+$2154 \citep{Zhong_2020}, and \unit[3.8]{kpc} for 1~RXS~J170849 \citep{Olausen_2014}. \cite{Atel_13553} estimates a distance of 4.8 to \unit[8.1]{kpc} for Swift J1818.0$-$1607, and we optimistically perform our analysis assuming a distance of \unit[4.8]{kpc}. \newline *Burst 2651 occurred 87 \si{\s} before Virgo was in observation mode. \newline $^{\dagger}$\url{https://heasarc.gsfc.nasa.gov/W3Browse/fermi/fermigtrig.html}}
\end{center}
\end{table*}

We follow \cite{Abbott_2019} to search for short-duration signals (as potentially emitted by \acp{fmode}) and long-duration signals (such as might accompany observed \acp{QPO}).
Each search combines the data from two (or more when available in the case of the short-duration search) detectors into a \acused{tf}\ac{tf} map and then forms groups of pixels, called clusters, to search for gravitational-wave signals.
The three electromagnetic bursts without a known source are only analyzed with the short-duration search.
The searches are described in the following sections.

\subsection{Short-duration search}\label{subsec:short-method}

The search for short-duration transient \acp{GW} (milliseconds to seconds) is motivated by a potential signal associated with \ac{fmode} oscillations in the magnetar's core. Because the frequencies of the expected gravitational-wave signals can be as high as several kilohertz~\citep{gw170817fmodes,nsfmodes}, the search ranges in frequency from 50 Hz to 4000 Hz, extending to higher frequencies than the other unmodeled searches (most notably LIGO-Virgo-KAGRA's burst searches associated with gamma-ray bursts \citep{2021ApJ...915...86A, LIGO_o3b_GRB} and fast radio bursts \citep{o3a_FRB}.

Data are searched with X-pipeline, which is an unmodeled, coherent search pipeline \citep{Sutton2010NJPh,2012PhRvD..86b2003W}. The X-pipeline algorithm coherently combines the data from each detector in the network to produce a multi-resolution \ac{tf} map displaying the energy in each pixel. The brightest $1\%$ of these pixels are then selected, and neighboring bright pixels are combined into clusters, which are then assigned a ranking statistic. A large fraction of background clusters are rejected applying vetos built from the coherent and incoherent power across the detector network.
Other details on the parameters of the short-duration search that have been modified are summarized in \appendixref{app:short-duration-parameters}.

The search for short-duration \acp{GW} is comprised of two components: an \unit[8]{s} duration on-source window centered on the magnetar burst time (a \textit{centered on-source} window) so as to optimize sensitivity at the time when \ac{GW} emission is most probable, and a \unit[500]{s} long on-source window beginning just after the centered on-source window (a \textit{delayed on-source} window). The longer delayed on-source window is meant to search for short-duration signals emitted during the time following the burst, analogous to the \acp{QPO} that have been observed in the giant flares.
For both on-source windows, \unit[3]{h} of data taken symmetrically about the burst time with an extra gap of \unit[16]{s} before and after the on-source window are used to determine the background significance. 

The two short-duration search components are considered independent, and no trials factor will be incurred between the searches.
However, since the delayed on-source window is \unit[500]{s} long and starts just after the end of the centered on-source window, it overlaps with background data for the centered on-source component of the search.
This introduces the possibility of there being a signal detected in the delayed on-source window, and this signal being included in the background of the centered on-source window.
We mitigate this possibility by examining the results of the delayed on-source search and verifying the absence of such a signal before viewing the results of the centered on-source window.
A summary of the main characteristics of both searches is reported in \tableref{tab:search_summary}.

\subsection{Long-duration search}\label{subsec:long-method}

Following previous searches \citep{quitzowjames2016thesis, Quitzow_James_2017, Abbott_2019}, the \ac{O3} long-duration signal search is carried out with the \ac{STAMP} pipeline \citep{PhysRevD.83.083004}. \ac{SNR} \ac{tf} maps are built using the cross-power between two detectors.
We then use \ac{STAMP}'s seedless clustering algorithm, which generates clusters with quadratic B\'{e}zier curves \citep{PhysRevD.88.083010}, to search for \ac{GW} signals. 
Restricting the variation of the clusters in frequency to a maximum of 10\% allows us to target 
nearly-monochromatic \acp{GW} signals potentially emitted from the mechanisms responsible for \acp{QPO} while reducing computational resources \citep{Abbott_2019}. Clusters are ranked according to their \ac{SNR} defined as the sum of the pixels' SNR that compose the cluster \citep{PhysRevD.88.083010}.

The on-source window starts \unit[4]{s} before the burst time and ends \unit[1600]{s} after.
Several on-source windows had data removed from the analysis, as discussed in \appendixref{app:long-duration-data}.
The frequency range is $24\hz$ -- $2500\hz$, limited by seismic noise at low frequencies, and going to just above the highest observed \ac{QPO} in the tail of the 2004 giant flare~\citep{mereghetti05}. Data from \ac{LHO} and \ac{LLO} are used when available, with data from Virgo used for bursts that also have data from either \ac{LHO} or \ac{LLO} (see \tableref{tab:table_triggers} for available detectors for each burst).

The background for each burst is estimated using \unit[58,800]{s} of off-source data with both detectors in observing mode which is as close as possible to the burst's on-source window while excluding the on-source windows of all of the bursts.
This data is broken up into 36 off-source analysis windows,
and pairing data from each detector from analysis windows with non-overlapping times allows us to create 1260 ($36^2 - 36$) \ac{SNR} \ac{tf} maps (which we will refer to as background segments) to calculate a background distribution for each burst.

\renewcommand{\arraystretch}{1.5}

\begin{table*}
\begin{center}
\footnotesize
\begin{tabular}{cccccc}
	\hline

	      Search name & Pipeline & \thead{On-source \\ window [s]}  & \thead{On-source \\interval [s]}  & \thead{Off-source \\ window [s]}& \thead{Frequency \\range [Hz]}  \\[4pt]

	\hline
	\hline
	      Centered on-source   & X pipeline & \thead{8} & \thead{$[-4,+4]$} & \thead{10,800} & \thead{$50 - 4000$} \\[4pt]
	      Delayed on-source  &  X pipeline & \thead{500} & \thead{$[+4, +504]$} & \thead{10,800} & \thead{$50 - 4000$} \\[4pt]
	     Long-duration & STAMP & \thead{1604} & \thead{$[-4, +1600]$} & \thead{58,800} & \thead{$24 - 2500$} \\[4pt]
                        
	\hline
\end{tabular}
	\caption{\label{tab:search_summary} On-source, off-source and frequency range of searches performed. The on-source interval is centered around the magnetar burst time.}
\end{center}
\end{table*}

\section{Results}\label{sec:results}

The results of the short-duration and long-duration searches for each burst are presented in the following sections. Clusters found in the on-source windows of the short-duration search are ranked by their p-value, which is the probability of having a cluster as significant or more in the on-source window under the null hypothesis. They are calculated considering the clusters in the background with a ranking statistic larger than the on-source cluster.
We characterize the significance of the on-source clusters of the long-duration search with the \ac{FBS}
\begin{equation}
  \mathrm{FBS} = N_\geq/N_\mathrm{Total}\;,
\end{equation}
where $N_\geq$ is the number of clusters in the respective burst's background with an \ac{SNR} greater than or equal to the on-source cluster \ac{SNR} and $N_\mathrm{Total}$ is the total number of clusters in that background.

To estimate the search sensitivity for each burst, we inject waveforms in the data and calculate the amplitude for which 50\% of the signals have a detection statistic equal to or greater than the respective on-source most significant cluster statistic.
We then derive the \ac{hrss} and \ac{EGW} for this amplitude at 50\% detection efficiency ($h_\mathrm{rss}^{50\%}$ and $E_\mathrm{GW}^{50\%}$).
The definition of \ac{hrss} is
\begin{equation}\label{hrss_eq}
  \ac{hrss} = \sqrt{\definiteIntegral{-\infty}{\infty}{\left|h_+(t)\right|^2 + \left|h_\times(t)\right|^2}{t}}\;,
\end{equation}
where $h_+(t)$ and $h_\times(t)$ are the two signal polarizations. In \appendixref{app:injections} we derive \ac{hrss} and \ac{EGW} for different waveforms used in the analysis.

When calculating \ac{EGW}, we use \unit[9]{kpc} for SGR 1935$+$2154 \citep{Zhong_2020}.
The distance of Swift J1818.0$-$1607 is estimated to be in the range of 4.8 to \unit[8.1]{kpc} \citep{Atel_13553}; we calculate \ac{EGW} with the optimistic closer value of \unit[4.8]{kpc}.
For the magnetar bursts with unknown source(s), we consider 1~RXS~J170849 as the source and use \unit[3.8]{kpc} for the distance \citep{Olausen_2014}.

\renewcommand{\arraystretch}{1}

\begin{table}[ht]

\begin{center}

\begin{tabular}{| c c c c | }

  \hline

  & Long-Duration & Centered (8 \si{\s}) & Delayed (500 \si{\s}) \\
  Burst & \ac{FBS} & p-value & p-value \\

  \hline\hline

  2651 & $9.1\times 10^{-1}$ &-&-  \\
  \hline

  2652 & $2.1\times 10^{-1}$   & $1.0$  & $4.1\times 10^{-1}$    \\
  \hline

  2653 & $7.7\times 10^{-1}$   & $1.8\times 10^{-2}$ & $3.5\times 10^{-3}$    \\
  \hline

  2654 & $8.5\times 10^{-1}$   & $9.9\times 10^{-2}$  & $6.6\times 10^{-1}$   \\
  \hline

  2655 & $7.8\times 10^{-1}$   & $1.0$  & $7.3\times 10^{-1}$    \\
  \hline

  2656 & $9.7\times 10^{-1}$   & $8.6\times 10^{-3}$  & $5.2\times 10^{-1}$    \\
  \hline

  2657 & $1.6\times 10^{-1}$   & $6.7\times 10^{-2}$  & $4.8\times 10^{-1}$    \\
  \hline

  2660 & $2.2\times 10^{-2}$   &  $4.8\times 10^{-1}$  & $4.9\times 10^{-1}$    \\
  \hline

  2661 & $9.0\times 10^{-1}$   & $1.9\times 10^{-1}$  & $7.4\times 10^{-1}$    \\
  \hline

  2665 & $8.2\times 10^{-1}$   &-&-  \\
  \hline

  2668 & $9.9\times 10^{-1}$   & $1.0$  & $4.3\times 10^{-1}$   \\
  \hline

  2669 & - & $1.0\times 10^{0}$  & $6.3\times 10^{-1}$    \\
  \hline

  2670 & - & $ 2.9\times 10^{-1}$  & $7.8\times 10^{-1}$    \\
  \hline

  2671 & - & $3.4\times 10^{-1}$  &-  \\
  \hline

  2673 & $6.8\times 10^{-1}$   & $1.0$  & $4.8\times 10^{-1}$    \\
  \hline

  2674 & $5.9\times 10^{-1}$   & $1.0$  &$5.0\times 10^{-2}$    \\
  \hline

\end{tabular}

  \caption{\label{tab:open_box_results} False alarm statistic for the most significant cluster found for each burst by all searches. \ac{FBS} is reported for the long-duration search, and the p-value for the short-duration searches. A p-value of 1 indicates that there were no clusters in the on-source window that survived the veto cuts.
  Table entries of `-' indicate that no value is recorded, because no search was run.
  Bursts 2669-2671 are from the magnetar with no identified source, and the long-duration search was not conducted on these.
  Bursts 2651 and 2665 were missing segments of data from the on-source window such that the short-duration searches could not run, and burst 2671 had data missing in the delayed on-source window such that the delayed on-source search could not be run.}

\end{center}

\end{table}

\subsection{Short-duration search results and upper limits}\label{subsec:short_upper_limits}

\begin{figure}[h!]
\centering
{\includegraphics[width=.7\textwidth]{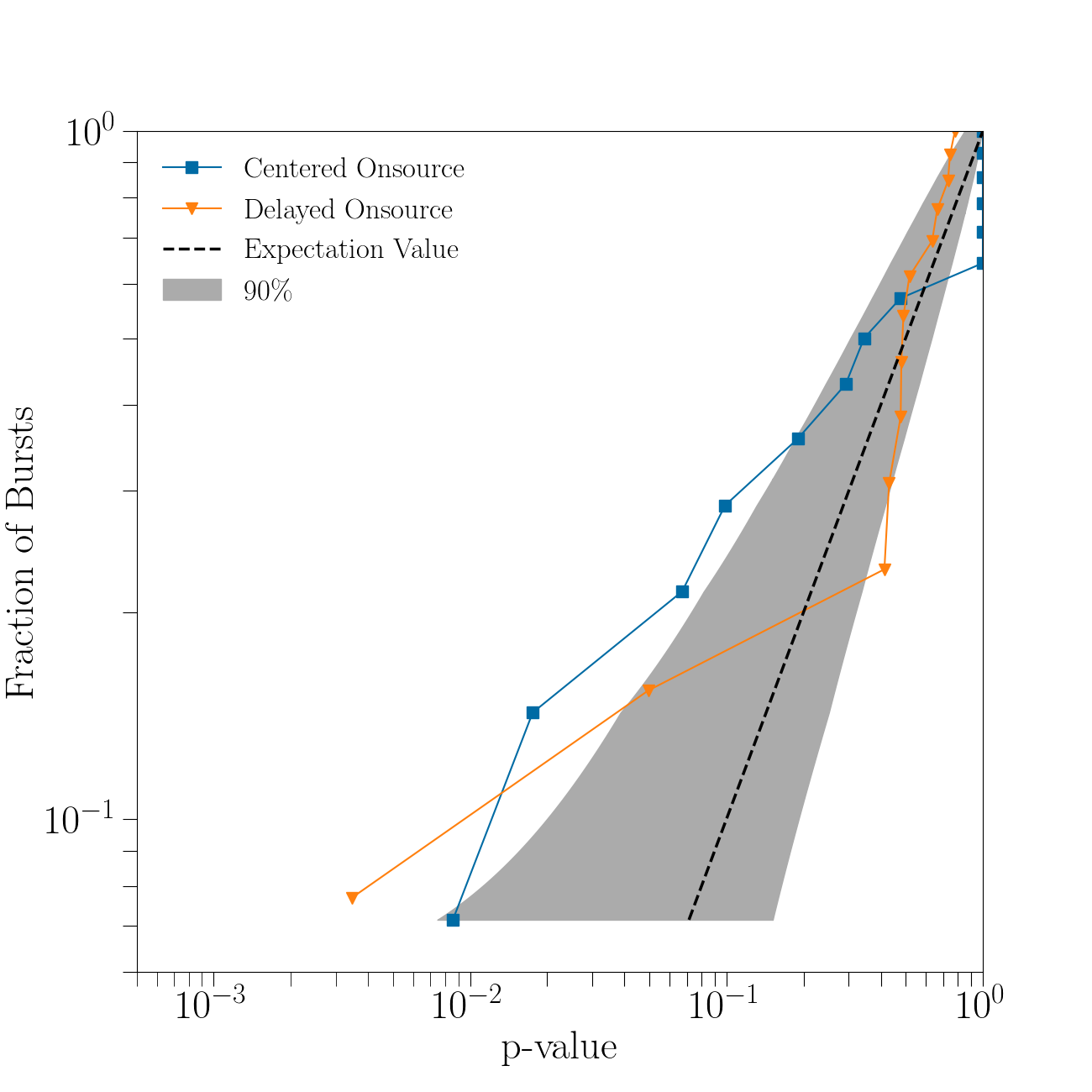}}
\caption{
Cumulative distribution of the p-values of the most significant clusters found by the short-duration searches. The expectation value and 90\% band are calculated as in \citep{o3a_FRB}; the expectation value is calculated assuming a uniform distribution of p-values (given by the null hypothesis), with the corresponding 90\% confidence band shown by the shaded region.  The lowest p-value in the delayed on-source search is $3.5 \times 10^{-3}$ from burst 2653, and is determined to be most likely an instrumental artifact through arguments invoking both astrophysics and the characteristics of the detector.  Several centered on-source search clusters fall outside of the $90\%$ band, but many of them have properties inconsistent with what one would expect from an astrophysical source, so this is likely a result of small number statistics. Although the most significant of these, which appears in burst 2656, has p-value $8.6 \times 10^{-3}$ and a peak frequency (\unit[1577]{Hz}) which is consistent with expectations for an f-mode, we provide arguments for why this is not the case in \subsectionref{subsec:short_upper_limits}.
}
\label{fig:p-values}
\end{figure}

All bursts listed in \tableref{tab:table_triggers} have been considered by the short-duration search except bursts 2651 and 2665 because of missing data in the on-source windows. For all the others, the short-duration centered and delayed on-source searches have found clusters whose significance is estimated in terms of their p-value reported in \tableref{tab:open_box_results}. The cumulative distribution of these p-values is represented in \figureref{fig:p-values}.

In the short-duration searches simulated signals are added into the on-source data and time-frequency maps are processed similarly to the on-source time-frequency maps.
The injected signals are chosen such that they adequately model a short-duration transient consistent with a magnetar \ac{fmode} signal, and also such that they cover a reasonable range of frequencies outside of the \ac{fmode} frequency range.  We inject sine-Gaussian waveforms, described by \equationmention{1} of \citep{2017ApJ...841...89A}, at a range of frequencies between \unit[70]{Hz} and \unit[3560]{Hz}. We also inject a series of ringdown waveforms characterized by damped sinusoids, where the damping term is exponentially decaying and frequencies are between \unit[1500]{Hz} and \unit[2020]{Hz}. We also include a series of white noise burst signals ranging in frequency from \unit[100]{Hz} to \unit[1000]{Hz} to probe search sensitivity at lower frequencies.
For each injected waveform we give the lowest value of the \hrssulF{}, and the corresponding \egwulF{} considering all bursts from SGR 1935$+$2154, Swift J1818.0$-$1607, and the unknown source in \tableref{tab:short_duration_injections_1935_50p_ULboxes}, \tableref{tab:short_duration_injections_1818_50p_ULboxes}, and \tableref{tab:short_duration_injections_unknown_50p_ULboxes}, respectively. In \tableref{tab:8s_fmodeUL} we give the results for two waveforms that best model the \ac{fmode} for each burst.\\

We now discuss the most significant clusters found for bursts 2653 and 2656 by the short-duration search.

\begin{figure}[ht]
\centering

\gridline{\fig{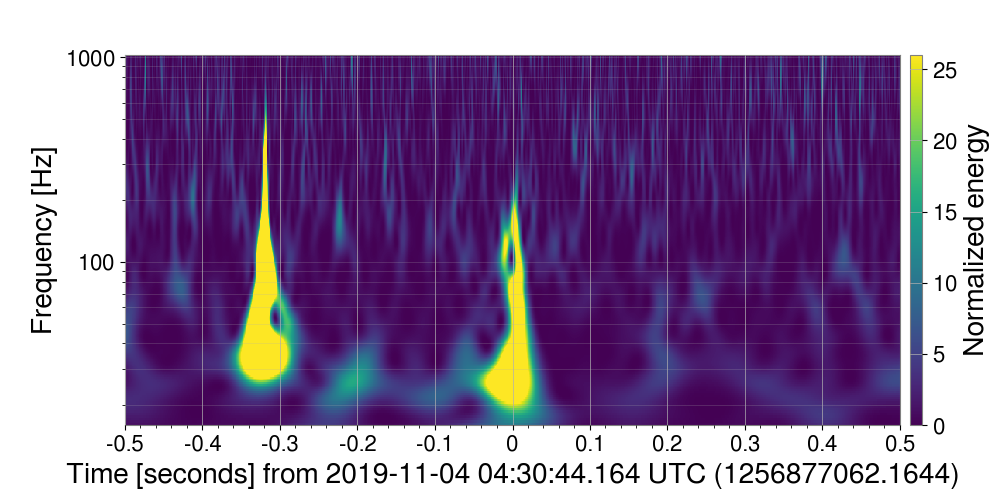}{0.5\textwidth}{}
\fig{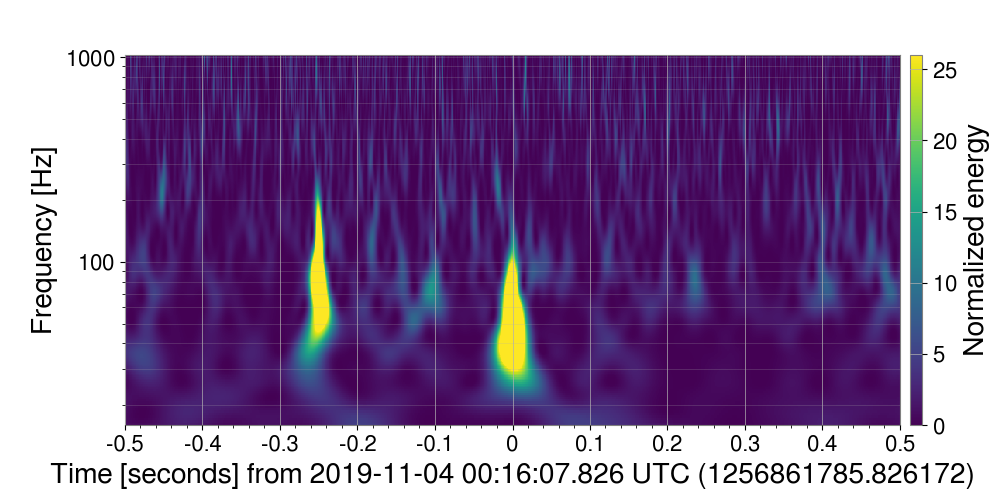}{0.5\textwidth}{}}

\caption{Spectrograms of the \ac{LHO} data (left) at the time of the most significant cluster found in the on-source window of burst 2653, and an instrumental artifact appearing also in \ac{LHO} data (right). The time separation between these two events is $\sim$ \unit[15,277]{s}, which is larger than the size of the background window in our analysis, and therefore the instrumental artifact was not included in the background. These two spectrograms display very similar structure, with a similar double short-duration transients separated in both cases by $\sim$ \unit[0.3]{s}. Several instrumental artifacts of this nature were found by Omicron \citep{robinet2020omicron} within the day of burst 2653. The second most significant cluster in the same delayed on-source analysis of burst 2653 also has this same double-peaked structure.}

\label{fig:glitch_H1}
\end{figure}

\textbf{Burst 2653}:
Two significant clusters are detected by the delayed on-source search.
The most significant cluster has a p-value of $3.5 \times 10^{-3}$, a peak frequency of \unit[97]{Hz} and duration of \unit[31]{ms}. This is the outlier displayed in \figureref{fig:glitch_H1}. X-pipeline identifies another loud cluster \unit[50]{s} earlier with the same p-value, a peak frequency at \unit[228]{Hz} and a duration of \unit[7.8]{ms}.
Given their frequency and duration, each of these clusters is only a couple of cycles long.
Both clusters display similar characteristics in the \ac{LHO} data, appearing in spectrograms as a short-duration and low-frequency spike. In both cases, another high-SNR spike is seen in the data $\sim$ \unit[0.3]{s} from the time of the cluster as shown on \figureref{fig:glitch_H1}. Several other of these high-SNR features are detected on the same day as trigger 2653, and these are also in groups of two spaced out by $~.3$ \si{\s}. While X-pipeline does not reconstruct either of these neighboring spikes in a coherent cluster, their presence at such a constant and short time-separation from the cluster strongly suggests that each of these clusters has a terrestrial origin. 
While X-pipeline conducts a coherent search, there remains poor coherence across the two detectors network for these clusters. When the energy per time-frequency pixel is standardized to be $1$ unit for Gaussian data, this signal was detected with $165.3$ units in \ac{LHO}, and only $6.2$ units and $12.8$ units of energy in \ac{LLO} and Virgo respectively. The ratio between the \ac{LHO} and \ac{LLO} energies is more than can be accounted for by the ratio of antenna factors squared (4.8 for \ac{LHO} - \ac{LLO}, and 1.3 \ac{LHO} - Virgo) at the time of the burst.
Finally, neither of these clusters is visibly identifiable in \ac{LLO} nor Virgo time-frequency maps, meaning they are likely instrumental artifacts in \ac{LHO} only.

\textbf{Burst 2656}: the most significant cluster found by the centered on-source search has a p-value of $8.6 \times 10^{-3}$. This cluster is \unit[63]{ms} long, and its frequency lies in \unit[$1560 - 1608$]{Hz} which matches the expected frequency range of neutron star \ac{fmode} oscillations.
The cluster appears $\sim$ \unit[3.1]{s} before the burst time. There are no known physical mechanisms to produce \acp{GW} so long before the electromagnetic emission; while this does not rule out this cluster being astrophysical, it makes it less plausible.
\par
\begin{figure}[ht]
\centering
\gridline{\fig{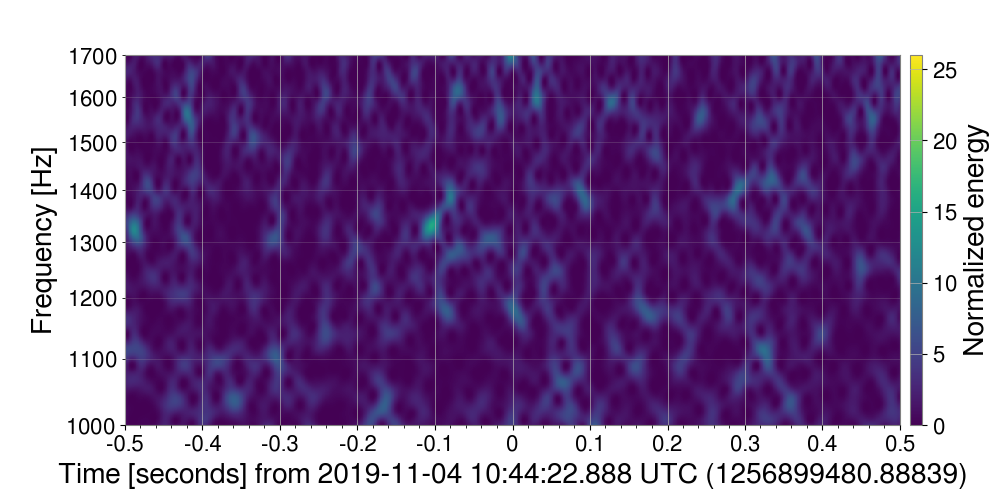}{0.5\textwidth}{}
  \fig{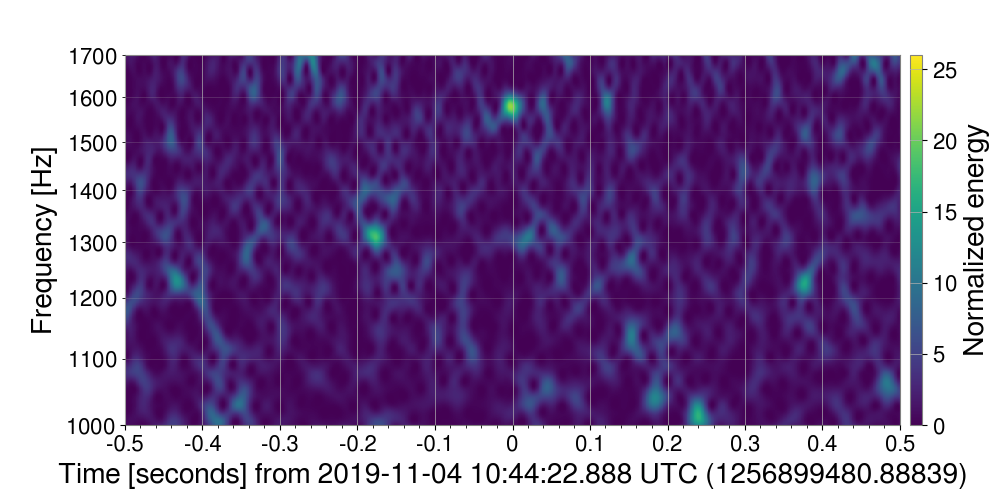}{0.5\textwidth}{}
	}
\caption{Spectrograms of the \ac{LHO} (left) and \ac{LLO} (right) data around the time of the most significant cluster found by the centered-window short-duration search of burst 2656. Even though the cluster is only barely visible in \ac{LLO} data, these spectrograms show that the ambient noise in each detector is in a normal state around the time of the flare.}%
\label{fig:2656_H1_Q}
\end{figure}

 Unlike the clusters found for burst 2653, this one is only visible in the \ac{LLO} data spectrogram shown on \figureref{fig:2656_H1_Q} that also illustrates that no high-SNR glitches are present in the ambient noise at the time of the burst. The data quality monitor tool iDQ \citep{essick2020idq} indicates that the probability of this cluster being an instrumental artifact in \ac{LLO} is $62\%$, and $23\%$ in \ac{LHO}.

The number of analyses in the long duration and the short-duration searches (40 in total across the three on-source windows of the 16 bursts) imply this cluster has a $\approx 29\%$ chance of being background noise.
This, and the high probability that the cluster is due to an instrumental artifact, imply it is highly unlikely to be a signal of astrophysical origin. Nevertheless, we can calculate the strain and gravitational-wave energy required for such a signal assuming it is of astrophysical origin. Using the ringdown waveform with 100 ms damping time and 1590 Hz frequency, the $h_{rss}$ and $E_{\rm GW}$ estimated at 50\% detection efficiency are $2.2\times10^{-22}\,{\rm Hz}^{-1/2}$ and $1\times10^{47}\,{\rm erg}$, respectively. For this burst, this implies the ratio of gravitational-wave to electromagnetic energy is $\Eratioflat=6\times10^6$. It is difficult to imagine a physical scenario whereby this much more energy is deposited into gravitational rather than electromagnetic waves, adding further weight to the conclusion that this cluster is not of astrophysical origin.

\subsection{Long-duration search results and upper limits}\label{subsec:long_upper_limits}

\begin{figure}[ht]
\centering
	\gridline{
          \fig{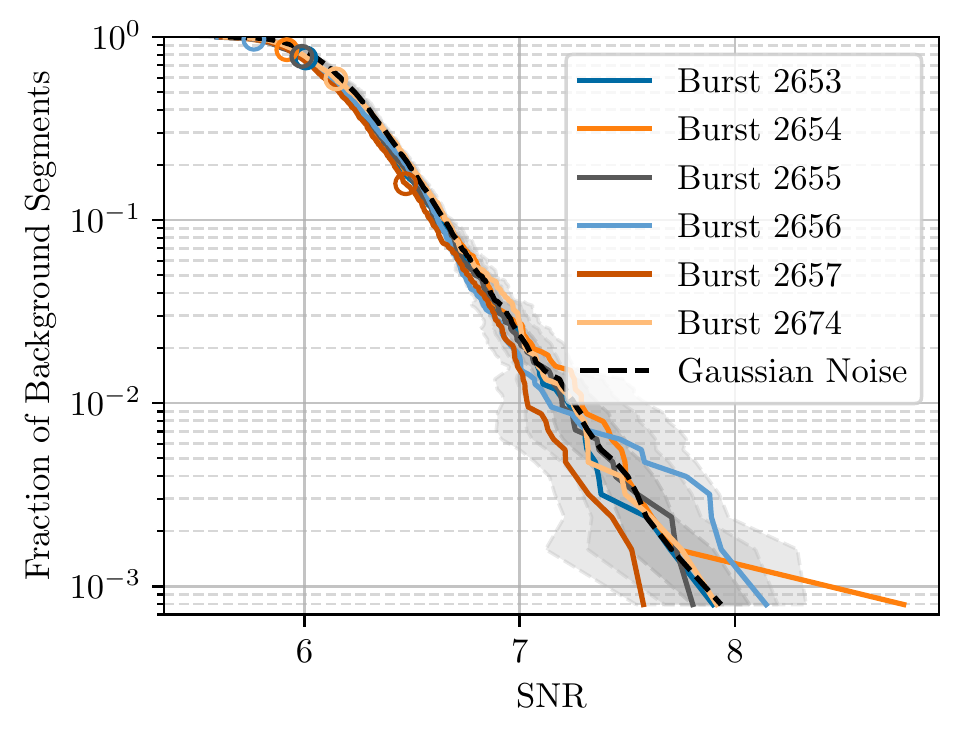}{0.45\textwidth}{(\ac{LHO} and \ac{LLO})}
          \fig{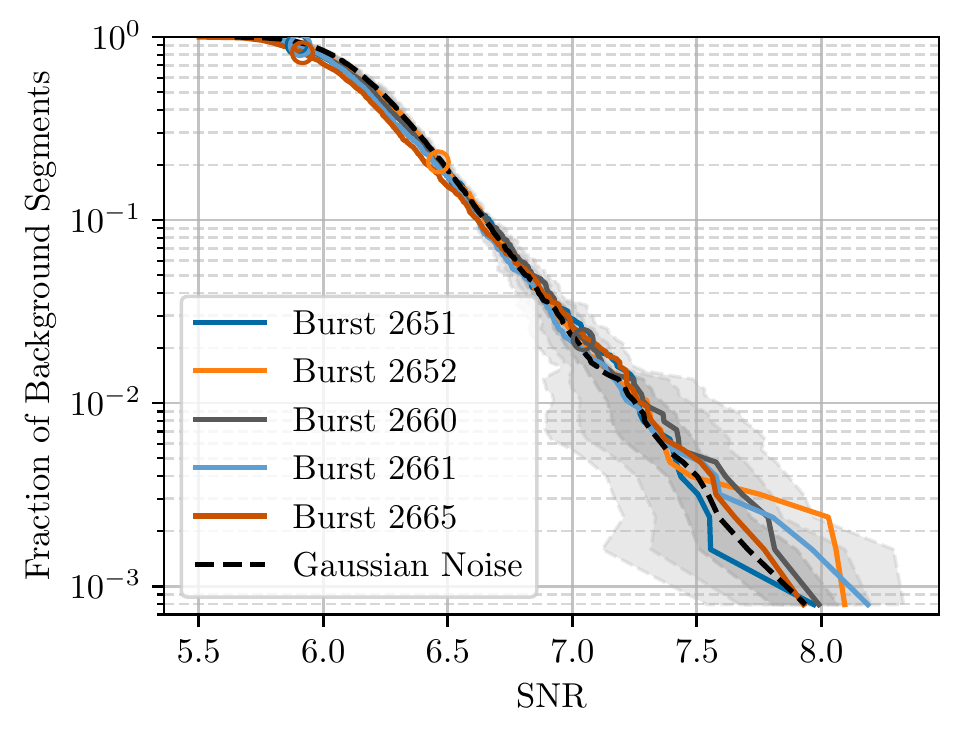}{0.45\textwidth} {(\ac{LHO} and Virgo)}
	  }
	\gridline{
          \fig{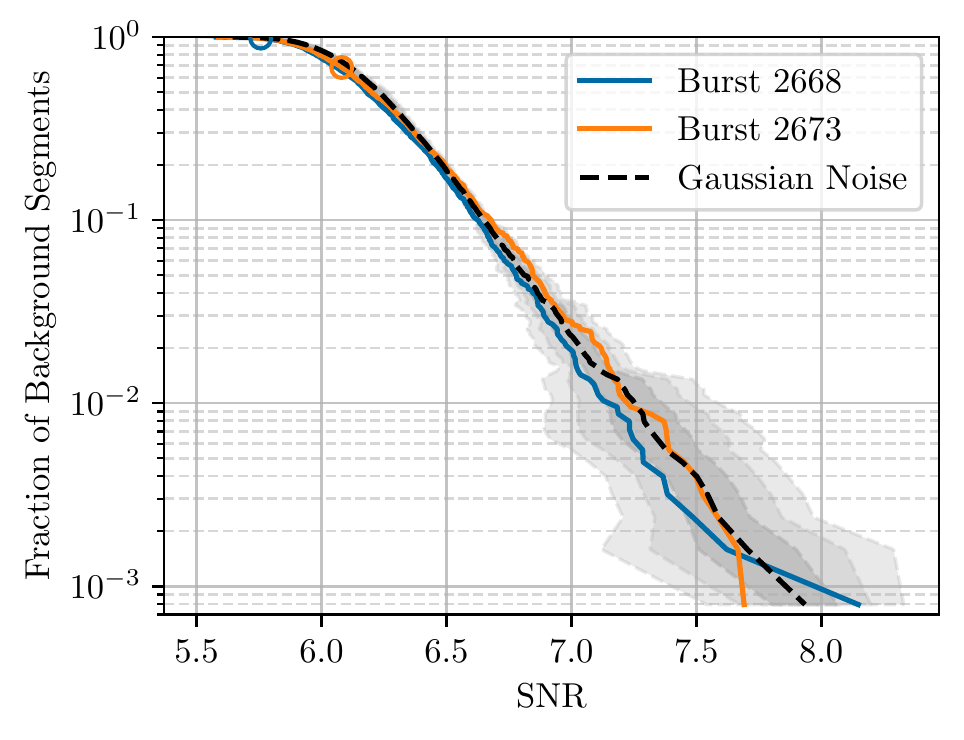}{0.45\textwidth} {(\ac{LLO} and Virgo)}
          }
\caption{
  Distributions of background and on-source clusters for each burst analyzed with data from \ac{LHO}, \ac{LLO} and Virgo. The background of each burst is displayed along with the SNR of its on-source cluster denoted by a circle. The dashed black line is the mean, and the grey contours are the one, two, and three standard deviations of the distribution of FBS obtained simulating Gaussian noise colored with the \ac{aLIGO} design sensitivity curve \citep{aLIGO_design}.
}
\label{fig:long-backgrounds-all}
\end{figure}

All bursts from \tableref{tab:table_triggers} have been analyzed by the long-duration search except the 3 bursts from the unknown source. The on-source window results for each burst are shown in \tableref{tab:open_box_results}, which lists the \ac{FBS} of the most significant cluster for each burst, determined using the background clusters. No interesting cluster has been found as the most significant cluster has an \ac{FBS} of $0.02$. \figureref{fig:long-backgrounds-all} compares the most significant on-source cluster to the corresponding background distribution for each burst.

\begin{figure}[h!]
{\includegraphics[width=\linewidth]{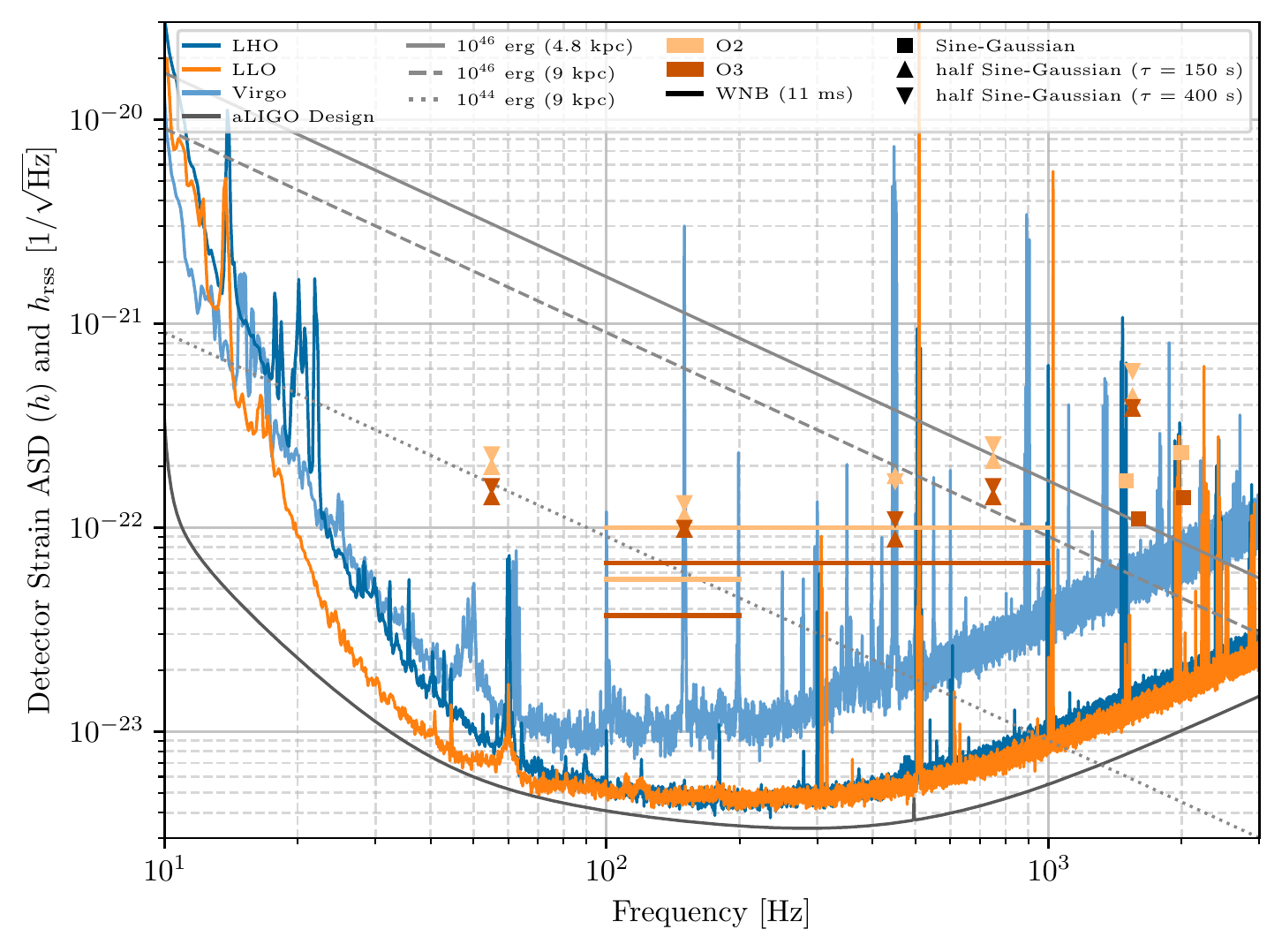}}
\caption{Upper limits obtained by the short-duration and long-duration searches. The most constrained \hrssulF{} across all bursts with known sources are displayed and compared to O2 results. The detectors' amplitude strain density curves correspond to the best sensitivity for each detector during O3 \citep{O3LIGOsensitivity, O3VirgoSensitivity, Buikema2020}.}
\label{fig:hrssestimate}
\end{figure}

As in the \ac{O2} search \citep{Abbott_2019}, two families of waveforms, half sine-Gaussians and exponentially decaying sinusoids (ringdowns), are injected at five frequencies (\unit[55]{Hz}, \unit[150]{Hz}, \unit[450]{Hz}, \unit[750]{Hz}, and \unit[1550]{Hz}) and two damping times (\unit[150]{s} and \unit[400]{s}).
The best upper limit in terms of \hrssulF{} is obtained for burst 2656, identified with SGR 1935$+$2154, which has the largest sum squared antenna factors.
The results for burst 2656 indicate an increase in search sensitivity from \ac{O2} to \ac{O3} ranging from a factor of $\sim1.2$ to $\sim2.0$, which follows roughly the detectors' sensitivity increase, although this improvement also depends on additional considerations such as the detector antenna factors at the time of the burst.
The half sine-Gaussian \hrssulF{} for burst 2656 are plotted against the best sensitivity curves of \ac{LHO} and \ac{LLO} during O3 \citep{O3LIGOsensitivity} in \figureref{fig:hrssestimate}, along with \hrssulF{} from \ac{O2} as a comparison. In \tableref{tab:long_upper_limits_2656} and \tableref{tab:long_upper_limits_2674} we report \hrssulF and \egwulF for bursts 2656 and 2674, which provide the best upper limits among all bursts emitted by SGR 1935$+$2154 and Swift J1818.0$-$1607, respectively.

The optimistic closer distance estimate for Swift J1818.0$-$1607 of 4.8 \si{kpc} \citep{Atel_13553} is roughly half the distance of SGR 1935$+$2154, so although \hrssulF{} results are slightly worse, upper bounds on \egwulF{} are 2.6 times better, with the best results from burst 2674 of \unit[$2.8 \times 10^{43}$]{erg} at \unit[55]{Hz}. All upper limits for bursts 2656 and 2674 are presented in \tableref{tab:long_upper_limits_2656} and \tableref{tab:long_upper_limits_2674}, respectively.

\renewcommand{\arraystretch}{1}

\begin{table*}
\begin{center}
\begin{tabular}{| c c c c c c c c c | }
  \hline
  Injection Type & Frequency (\si{Hz}) & Duration/ & \multicolumn{3}{c}{Centered ($8$ \si{\s})} & \multicolumn{3}{c |}{Delayed ($500$ \si{\s})} \\ 
  & & $\tau$ (\si{\ms}) & Burst & \hrssulF{} ($\si{\Hz}^{-1/2}$) & \egwulF{} (\si{\erg}) & Burst & \hrssulF{} ($\si{\Hz}^{-1/2}$) & \egwulF{} (\si{\erg}) \\[0.5ex] 
  \hline\hline
 Sine Gaussian & 70 & 14.3 & 2655 & $7.3\times 10^{-23}$ & $3.2\times 10^{43}$ & 2656 & $8.5\times 10^{-23}$ & $4.4\times 10^{43}$ \\
 Sine Gaussian & 90 & 11.1 & 2668 & $5.8\times 10^{-23}$ & $3.3\times 10^{43}$ & 2656 & $7.4\times 10^{-23}$ & $5.5\times 10^{43}$ \\
 Sine Gaussian & 145 & 6.89 & 2668 & $5.7\times 10^{-23}$ & $8.4\times 10^{43}$ & 2656 & $6.6\times 10^{-23}$ & $1.1\times 10^{44}$ \\
 Sine Gaussian & 290 & 3.45 & 2655 & $5.8\times 10^{-23}$ & $3.5\times 10^{44}$ & 2656 & $6.7\times 10^{-23}$ & $4.7\times 10^{44}$ \\
 Sine Gaussian & 1100 & 0.909 & 2655 & $1.3\times 10^{-22}$ & $2.6\times 10^{46}$ & 2656 & $1.5\times 10^{-22}$ & $3.4\times 10^{46}$ \\
 Sine Gaussian & 1600 & 0.625 & 2655 & $1.1\times 10^{-22}$ & $3.8\times 10^{46}$ & 2656 & $1.3\times 10^{-22}$ & $5.2\times 10^{46}$ \\
 Sine Gaussian & 2020 & 0.495 & 2655 & $1.4\times 10^{-22}$ & $9.6\times 10^{46}$ & 2656 & $1.7\times 10^{-22}$ & $1.4\times 10^{47}$ \\
 Sine Gaussian & 2600 & 0.385 & 2655 & $3.3\times 10^{-22}$ & $9.2\times 10^{47}$ & 2656 & $4.0\times 10^{-22}$ & $1.3\times 10^{48}$ \\
 Sine Gaussian & 3100 & 0.323 & 2655 & $4.1\times 10^{-22}$ & $2.0\times 10^{48}$ & 2652 & $5.2\times 10^{-22}$ & $3.1\times 10^{48}$ \\
 Sine Gaussian & 3560 & 0.281 & 2655 & $5.1\times 10^{-22}$ & $4.0\times 10^{48}$ & 2652 & $6.2\times 10^{-22}$ & $6.0\times 10^{48}$ \\
 Sine Gaussian$^*$ & 1600 & 0.625 & 2655 & $1.1\times 10^{-22}$ & $3.8\times 10^{46}$ & 2656 & $1.3\times 10^{-22}$ & $5.2\times 10^{46}$ \\
 Sine Gaussian$^*$ & 2020 & 0.495 & 2655 & $1.4\times 10^{-22}$ & $9.6\times 10^{46}$ & 2656 & $1.7\times 10^{-22}$ & $1.4\times 10^{47}$ \\
 Ringdown & 1590 & 100 & 2655 & $1.9\times 10^{-22}$ & $1.1\times 10^{47}$ & - & - & - \\
 Ringdown & 1590 & 200 & 2655 & $1.8\times 10^{-22}$ & $9.6\times 10^{46}$ & - & - & - \\
 Ringdown & 1500 & 100 & 2668 & $2.5\times 10^{-22}$ & $1.8\times 10^{47}$ & - & - & - \\
 Ringdown & 1500 & 200 & 2668 & $2.3\times 10^{-22}$ & $1.5\times 10^{47}$ & - & - & - \\
 Ringdown & 2020 & 100 & 2655 & $2.2\times 10^{-22}$ & $2.4\times 10^{47}$ & - & - & - \\
 Ringdown & 2020 & 200 & 2655 & $2.2\times 10^{-22}$ & $2.3\times 10^{47}$ & - & - & - \\
 WNB & 100-200 & 11 & 2668 & $3.8\times 10^{-23}$ & $4.0\times 10^{43}$ & 2656 & $4.1\times 10^{-23}$ & $4.7\times 10^{43}$ \\
 WNB & 100-200 & 100 & 2668 & $5.3\times 10^{-23}$ & $7.9\times 10^{43}$ & 2656 & $6.1\times 10^{-23}$ & $1.0\times 10^{44}$ \\
 WNB & 100-1000 & 11 & 2655 & $6.7\times 10^{-23}$ & $1.7\times 10^{45}$ & 2656 & $7.5\times 10^{-23}$ & $2.1\times 10^{45}$ \\
 WNB & 100-1000 & 100 & 2655 & $1.1\times 10^{-22}$ & $4.9\times 10^{45}$ & 2656 & $1.5\times 10^{-22}$ & $7.9\times 10^{45}$ \\

  \hline
\end{tabular}
  \caption{\label{tab:short_duration_injections_1935_50p_ULboxes} $\hrss$ and isotropic energy estimated at 50\% detection efficiency for the most sensitive burst from SGR 1935$+$2154 for each injected waveform in the centered and delayed on-source windows of the short-duration search. All waveforms are elliptically polarized, except those denoted by $^*$, which have circular polarization.}
\end{center}
\end{table*}

\renewcommand{\arraystretch}{1}

\begin{table*}
\begin{center}
\begin{tabular}{| c c c c c c c c c | }
  \hline
  Injection Type & Frequency (\si{Hz}) & Duration/ & \multicolumn{3}{c}{Centered ($8$ \si{\s})} & \multicolumn{3}{c |}{Delayed ($500$ \si{\s})} \\ 
  & & $\tau$ (\si{\ms}) & Burst & \hrssulF{} ($\si{\Hz}^{-1/2}$) & \egwulF{} (\si{\erg}) & Burst & \hrssulF{} ($\si{\Hz}^{-1/2}$) & \egwulF{} (\si{\erg}) \\[0.5ex] 
  \hline\hline
 Sine Gaussian & 70 & 14.3 & 2674 & $6.9\times 10^{-23}$ & $8.2\times 10^{42}$ & 2673 & $9.2\times 10^{-23}$ & $1.4\times 10^{43}$ \\
 Sine Gaussian & 90 & 11.1 & 2674 & $6.1\times 10^{-23}$ & $1.1\times 10^{43}$ & 2673 & $8.7\times 10^{-23}$ & $2.1\times 10^{43}$ \\
 Sine Gaussian & 145 & 6.89 & 2674 & $5.7\times 10^{-23}$ & $2.4\times 10^{43}$ & 2674 & $8.5\times 10^{-23}$ & $5.4\times 10^{43}$ \\
 Sine Gaussian & 290 & 3.45 & 2674 & $5.9\times 10^{-23}$ & $1.0\times 10^{44}$ & 2674 & $9.1\times 10^{-23}$ & $2.4\times 10^{44}$ \\
 Sine Gaussian & 1100 & 0.909 & 2674 & $1.3\times 10^{-22}$ & $7.5\times 10^{45}$ & 2674 & $2.0\times 10^{-22}$ & $1.7\times 10^{46}$ \\
 Sine Gaussian & 1600 & 0.625 & 2674 & $1.2\times 10^{-22}$ & $1.2\times 10^{46}$ & 2674 & $1.7\times 10^{-22}$ & $2.6\times 10^{46}$ \\
 Sine Gaussian & 2020 & 0.495 & 2674 & $1.5\times 10^{-22}$ & $3.2\times 10^{46}$ & 2674 & $2.1\times 10^{-22}$ & $6.3\times 10^{46}$ \\
 Sine Gaussian & 2600 & 0.385 & 2674 & $3.5\times 10^{-22}$ & $2.9\times 10^{47}$ & 2674 & $4.9\times 10^{-22}$ & $5.7\times 10^{47}$ \\
 Sine Gaussian & 3100 & 0.323 & 2674 & $4.4\times 10^{-22}$ & $6.4\times 10^{47}$ & 2674 & $6.1\times 10^{-22}$ & $1.2\times 10^{48}$ \\
 Sine Gaussian & 3560 & 0.281 & 2674 & $5.3\times 10^{-22}$ & $1.2\times 10^{48}$ & 2674 & $7.1\times 10^{-22}$ & $2.2\times 10^{48}$ \\
 Sine Gaussian$^*$ & 1600 & 0.625 & 2674 & $1.2\times 10^{-22}$ & $1.2\times 10^{46}$ & 2674 & $1.7\times 10^{-22}$ & $2.6\times 10^{46}$ \\
 Sine Gaussian$^*$ & 2020 & 0.495 & 2674 & $1.5\times 10^{-22}$ & $3.2\times 10^{46}$ & 2674 & $2.1\times 10^{-22}$ & $6.3\times 10^{46}$ \\
 Ringdown & 1590 & 100 & 2674 & $2.1\times 10^{-22}$ & $3.8\times 10^{46}$ & - & - & - \\
 Ringdown & 1590 & 200 & 2674 & $2.1\times 10^{-22}$ & $4.0\times 10^{46}$ & - & - & - \\
 Ringdown & 1500 & 100 & 2673 & $2.6\times 10^{-22}$ & $5.3\times 10^{46}$ & - & - & - \\
 Ringdown & 1500 & 200 & 2673 & $2.6\times 10^{-22}$ & $5.3\times 10^{46}$ & - & - & - \\
 Ringdown & 2020 & 100 & 2674 & $2.6\times 10^{-22}$ & $9.3\times 10^{46}$ & - & - & - \\
 Ringdown & 2020 & 200 & 2674 & $2.5\times 10^{-22}$ & $8.7\times 10^{46}$ & - & - & - \\
 WNB & 100-200 & 11 & 2674 & $3.7\times 10^{-23}$ & $1.1\times 10^{43}$ & 2673 & $5.6\times 10^{-23}$ & $2.5\times 10^{43}$ \\
 WNB & 100-200 & 100 & 2674 & $5.3\times 10^{-23}$ & $2.2\times 10^{43}$ & 2673 & $7.8\times 10^{-23}$ & $4.8\times 10^{43}$ \\
 WNB & 100-1000 & 11 & 2674 & $7.0\times 10^{-23}$ & $5.1\times 10^{44}$ & 2674 & $1.0\times 10^{-22}$ & $1.1\times 10^{45}$ \\
 WNB & 100-1000 & 100 & 2674 & $1.2\times 10^{-22}$ & $1.5\times 10^{45}$ & 2674 & $1.8\times 10^{-22}$ & $3.6\times 10^{45}$ \\

  \hline
\end{tabular}
  \caption{\label{tab:short_duration_injections_1818_50p_ULboxes} $\hrss$ and isotropic energy estimated at 50\% detection efficiency for the most sensitive burst from SGR 1818$-$1607 for each injected waveform in the centered and delayed on-source windows of the short-duration search. All waveforms are elliptically polarized, except those denoted by $^*$, which have circular polarization.}
\end{center}
\end{table*}

\renewcommand{\arraystretch}{1}

\begin{table*}
\begin{center}
\begin{tabular}{| c c c c c c c c c | }
  \hline
  Injection Type & Frequency (\si{Hz}) & Duration/ & \multicolumn{3}{c}{Centered ($8$ \si{\s})} & \multicolumn{3}{c |}{Delayed ($500$ \si{\s})} \\ 
  & & $\tau$ (\si{\ms}) & Burst & \hrssulF{} ($\si{\Hz}^{-1/2}$) & \egwulF{} (\si{\erg}) & Burst & \hrssulF{} ($\si{\Hz}^{-1/2}$) & \egwulF{} (\si{\erg}) \\[0.5ex] 
  \hline\hline
 Sine Gaussian & 70 & 14.3 & 2669 & $4.2\times 10^{-23}$ & $1.9\times 10^{42}$ & 2669 & $5.0\times 10^{-23}$ & $2.7\times 10^{42}$ \\
 Sine Gaussian & 90 & 11.1 & 2669 & $4.0\times 10^{-23}$ & $2.8\times 10^{42}$ & 2669 & $4.6\times 10^{-23}$ & $3.8\times 10^{42}$ \\
 Sine Gaussian & 145 & 6.89 & 2669 & $3.3\times 10^{-23}$ & $5.1\times 10^{42}$ & 2669 & $3.9\times 10^{-23}$ & $7.0\times 10^{42}$ \\
 Sine Gaussian & 290 & 3.45 & 2669 & $3.6\times 10^{-23}$ & $2.4\times 10^{43}$ & 2669 & $4.1\times 10^{-23}$ & $3.1\times 10^{43}$ \\
 Sine Gaussian & 1100 & 0.909 & 2669 & $7.7\times 10^{-23}$ & $1.6\times 10^{45}$ & 2669 & $8.8\times 10^{-23}$ & $2.1\times 10^{45}$ \\
 Sine Gaussian & 1600 & 0.625 & 2669 & $6.2\times 10^{-23}$ & $2.2\times 10^{45}$ & 2669 & $7.7\times 10^{-23}$ & $3.3\times 10^{45}$ \\
 Sine Gaussian & 2020 & 0.495 & 2669 & $8.6\times 10^{-23}$ & $6.6\times 10^{45}$ & 2670 & $9.5\times 10^{-23}$ & $8.1\times 10^{45}$ \\
 Sine Gaussian & 2600 & 0.385 & 2669 & $2.0\times 10^{-22}$ & $5.9\times 10^{46}$ & 2669 & $2.4\times 10^{-22}$ & $8.8\times 10^{46}$ \\
 Sine Gaussian & 3100 & 0.323 & 2669 & $2.5\times 10^{-22}$ & $1.3\times 10^{47}$ & 2670 & $3.0\times 10^{-22}$ & $1.9\times 10^{47}$ \\
 Sine Gaussian & 3560 & 0.281 & 2669 & $2.9\times 10^{-22}$ & $2.3\times 10^{47}$ & 2670 & $4.0\times 10^{-22}$ & $4.5\times 10^{47}$ \\
 Sine Gaussian$^*$ & 1600 & 0.625 & 2669 & $6.2\times 10^{-23}$ & $2.2\times 10^{45}$ & 2669 & $7.7\times 10^{-23}$ & $3.3\times 10^{45}$ \\
 Sine Gaussian$^*$ & 2020 & 0.495 & 2669 & $8.6\times 10^{-23}$ & $6.6\times 10^{45}$ & 2670 & $9.5\times 10^{-23}$ & $8.1\times 10^{45}$ \\
 Ringdown & 1590 & 100 & 2670 & $1.1\times 10^{-22}$ & $6.8\times 10^{45}$ & - & - & - \\
 Ringdown & 1590 & 200 & 2670 & $1.1\times 10^{-22}$ & $7.0\times 10^{45}$ & - & - & - \\
 Ringdown & 1500 & 100 & 2669 & $1.8\times 10^{-22}$ & $1.6\times 10^{46}$ & - & - & - \\
 Ringdown & 1500 & 200 & 2671 & $2.7\times 10^{-22}$ & $3.6\times 10^{46}$ & - & - & - \\
 Ringdown & 2020 & 100 & 2670 & $1.3\times 10^{-22}$ & $1.6\times 10^{46}$ & - & - & - \\
 Ringdown & 2020 & 200 & 2670 & $1.3\times 10^{-22}$ & $1.6\times 10^{46}$ & - & - & - \\
 WNB & 100-200 & 11 & 2669 & $2.2\times 10^{-23}$ & $2.4\times 10^{42}$ & 2669 & $2.5\times 10^{-23}$ & $3.2\times 10^{42}$ \\
 WNB & 100-200 & 100 & 2669 & $3.2\times 10^{-23}$ & $5.0\times 10^{42}$ & 2670 & $3.6\times 10^{-23}$ & $6.5\times 10^{42}$ \\
 WNB & 100-1000 & 11 & 2669 & $3.9\times 10^{-23}$ & $9.8\times 10^{43}$ & 2669 & $4.6\times 10^{-23}$ & $1.4\times 10^{44}$ \\
 WNB & 100-1000 & 100 & 2669 & $7.1\times 10^{-23}$ & $3.4\times 10^{44}$ & 2669 & $8.2\times 10^{-23}$ & $4.5\times 10^{44}$ \\

  \hline
\end{tabular}
  \caption{\label{tab:short_duration_injections_unknown_50p_ULboxes} $\hrss$ and isotropic energy estimated at 50\% detection efficiency for the most sensitive burst from the unknown source for each injected waveform in the centered and delayed on-source windows of the short-duration search. All waveforms are elliptically polarized, except those denoted by $^*$, which have circular polarization.}
\end{center}
\end{table*}

\renewcommand{\arraystretch}{1}

\begin{table}[th]

\begin{center}

\begin{tabular}{| c c c c c c c c | }
  \hline
  Burst & Source & \multicolumn{3}{c}{Ringdown} & \multicolumn{3}{c |}{Sine Gaussian} \\ 
  & & \hrssulF{} ($\si{\Hz}^{-1/2}$) & \egwulF{} (\si{\erg}) & $\Eratio$ & \hrssulF{} ($\si{\Hz}^{-1/2}$) & \egwulF{} (\si{\erg}) & $\Eratio$ \\[1ex] 
  \hline\hline

  2652 & SGR 1935$+$2154 & $3.2\times 10^{-22}$ & $3.2\times 10^{47}$ & $2.2\times 10^{8}$ & $1.6\times 10^{-22}$ & $8.0\times 10^{46}$ & $5.6\times 10^{7}$ \\
  2653 & SGR 1935$+$2154 & $2.6\times 10^{-22}$ & $2.2\times 10^{47}$ & $1.9\times 10^{8}$ & $1.5\times 10^{-22}$ & $7.3\times 10^{46}$ & $6.4\times 10^{7}$ \\
  2654 & SGR 1935$+$2154 & $5.0\times 10^{-22}$ & $7.8\times 10^{47}$ & - & $2.9\times 10^{-22}$ & $2.7\times 10^{47}$ & - \\
  2655 & SGR 1935$+$2154 & $1.9\times 10^{-22}$ & $1.1\times 10^{47}$ & $1.9\times 10^{7}$ & $1.1\times 10^{-22}$ & $3.8\times 10^{46}$ & $6.6\times 10^{6}$ \\
  2656 & SGR 1935$+$2154 & $2.1\times 10^{-22}$ & $1.4\times 10^{47}$ & $6.4\times 10^{6}$ & $1.3\times 10^{-22}$ & $4.9\times 10^{46}$ & $2.2\times 10^{6}$ \\
  2657 & SGR 1935$+$2154 & $2.4\times 10^{-22}$ & $1.8\times 10^{47}$ & $6.6\times 10^{7}$ & $1.4\times 10^{-22}$ & $6.4\times 10^{46}$ & $2.3\times 10^{7}$ \\
  2660 & SGR 1935$+$2154 & $3.4\times 10^{-22}$ & $3.6\times 10^{47}$ & $3.0\times 10^{8}$ & $2.0\times 10^{-22}$ & $1.3\times 10^{47}$ & $1.1\times 10^{8}$ \\
  2661 & SGR 1935$+$2154 & $3.2\times 10^{-22}$ & $3.2\times 10^{47}$ & $2.4\times 10^{8}$ & $1.9\times 10^{-22}$ & $1.2\times 10^{47}$ & $9.0\times 10^{7}$ \\
  2668 & SGR 1935$+$2154 & $2.1\times 10^{-22}$ & $1.4\times 10^{47}$ & $1.8\times 10^{8}$ & $1.2\times 10^{-22}$ & $4.6\times 10^{46}$ & $6.0\times 10^{7}$ \\
  2669 & unknown & $1.1\times 10^{-22}$ & $7.3\times 10^{45}$ & - & $6.2\times 10^{-23}$ & $2.2\times 10^{45}$ & - \\
  2670 & unknown & $1.1\times 10^{-22}$ & $6.8\times 10^{45}$ & - & $6.6\times 10^{-23}$ & $2.5\times 10^{45}$ & - \\
  2671 & unknown & $2.0\times 10^{-22}$ & $2.2\times 10^{46}$ & - & $1.2\times 10^{-22}$ & $7.9\times 10^{45}$ & - \\
  2673 & Swift J1818$-$1607 & $2.7\times 10^{-22}$ & $6.6\times 10^{46}$ & - & $1.5\times 10^{-22}$ & $2.0\times 10^{46}$ & - \\
  2674 & Swift J1818$-$1607 & $2.1\times 10^{-22}$ & $3.8\times 10^{46}$ & - & $1.2\times 10^{-22}$ & $1.2\times 10^{46}$ & - \\

  \hline

\end{tabular}

  \caption{\label{tab:8s_fmodeUL} Upper limits on the $\hrss$ for 50\% detection efficiency and corresponding gravitational wave energy (\egwulF{}) for the centered on-source short-duration search. We present these values for the elliptically polarized $1600 \si{\Hz}$ Sine Gaussian waveform, and $100 \si{\ms}$ ringdown waveform at central frequency $1590 \si{\Hz}$. These are the specific waveforms of each type that best model the gravitational wave one would expect in conjuncture with an f-mode. The burst numbers are consistent with those used in the catalogue: http://www.ssl.berkeley.edu/ipn3/sgrlist.txt.}

\end{center}

\end{table}

\renewcommand{\arraystretch}{1}

\begin{table*}
\begin{center}
\begin{tabular}{| c c c c c c c c |}
  \hline
  Frequency (\si{\Hz}) & $\tau$ (\si{\s}) & \multicolumn{3}{c}{Half Sine-Gaussian} & \multicolumn{3}{c |}{Ringdown} \\
  & & Burst & \hrssulF{} ($\si{\Hz}^{-1/2}$) & \egwulF{} (\si{\erg}) & Burst & \hrssulF{} ($\si{\Hz}^{-1/2}$) & \egwulF{} (\si{\erg}) \\[0.5ex]
  \hline\hline
  55 & 150 & 2656 & $1.4 \times 10^{-22}$ & $7.3 \times 10^{43}$ & 2656 & $1.4 \times 10^{-22}$ & $7.8 \times 10^{43}$ \\
  - & 400 & 2656 & $1.6 \times 10^{-22}$ & $9.3 \times 10^{43}$ & 2656 & $1.7 \times 10^{-22}$ & $1.0 \times 10^{44}$ \\
  \hline
  150 & 150 & 2656 & $9.7 \times 10^{-23}$ & $2.6 \times 10^{44}$ & 2656 & $1.0 \times 10^{-22}$ & $2.9 \times 10^{44}$ \\
  - & 400 & 2656 & $1.0 \times 10^{-22}$ & $2.9 \times 10^{44}$ & 2656 & $1.1 \times 10^{-22}$ & $3.3 \times 10^{44}$ \\
  \hline
  450 & 150 & 2656 & $8.7 \times 10^{-23}$ & $1.9 \times 10^{45}$ & 2656 & $9.3 \times 10^{-23}$ & $2.1 \times 10^{45}$ \\
  - & 400 & 2656 & $1.1 \times 10^{-22}$ & $3.1 \times 10^{45}$ & 2656 & $1.2 \times 10^{-22}$ & $3.5 \times 10^{45}$ \\
  \hline
  750 & 150 & 2656 & $1.4 \times 10^{-22}$ & $1.4 \times 10^{46}$ & 2656 & $1.5 \times 10^{-22}$ & $1.7 \times 10^{46}$ \\
  - & 400 & 2656 & $1.6 \times 10^{-22}$ & $1.7 \times 10^{46}$ & 2656 & $1.7 \times 10^{-22}$ & $1.9 \times 10^{46}$ \\
  \hline
  1550 & 150 & 2656 & $3.8 \times 10^{-22}$ & $4.2 \times 10^{47}$ & 2656 & $3.8 \times 10^{-22}$ & $4.2 \times 10^{47}$ \\
  - & 400 & 2656 & $3.9 \times 10^{-22}$ & $4.4 \times 10^{47}$ & 2656 & $4.1 \times 10^{-22}$ & $5.0 \times 10^{47}$ \\
  \hline
\end{tabular}
  \caption{\label{tab:long_upper_limits_2656} Upper limits for \ac{hrss} and \ac{EGW} estimated at 50\% detection efficiency for half sine-Gaussian and ringdown waveforms for the most sensitive burst from SGR 1935+2154 for the long-duration search.}
\end{center}
\end{table*}

\renewcommand{\arraystretch}{1}

\begin{table*}
\begin{center}
\begin{tabular}{| c c c c c c c c |}
  \hline
  Frequency (\si{\Hz}) & $\tau$ (\si{\s}) & \multicolumn{3}{c}{Half Sine-Gaussian} & \multicolumn{3}{c |}{Ringdown} \\
  & & Burst & \hrssulF{} ($\si{\Hz}^{-1/2}$) & \egwulF{} (\si{\erg}) & Burst & \hrssulF{} ($\si{\Hz}^{-1/2}$) & \egwulF{} (\si{\erg}) \\[0.5ex]
  \hline\hline
  55 & 150 & 2674 & $1.6 \times 10^{-22}$ & $2.8 \times 10^{43}$ & 2674 & $1.7 \times 10^{-22}$ & $3.1 \times 10^{43}$ \\
  - & 400 & 2673 & $1.9 \times 10^{-22}$ & $3.7 \times 10^{43}$ & 2673 & $2.0 \times 10^{-22}$ & $4.1 \times 10^{43}$ \\
  \hline
  150 & 150 & 2674 & $1.3 \times 10^{-22}$ & $1.3 \times 10^{44}$ & 2674 & $1.4 \times 10^{-22}$ & $1.5 \times 10^{44}$ \\
  - & 400 & 2674 & $1.4 \times 10^{-22}$ & $1.5 \times 10^{44}$ & 2674 & $1.4 \times 10^{-22}$ & $1.6 \times 10^{44}$ \\
  \hline
  450 & 150 & 2674 & $1.2 \times 10^{-22}$ & $1.1 \times 10^{45}$ & 2674 & $1.3 \times 10^{-22}$ & $1.2 \times 10^{45}$ \\
  - & 400 & 2674 & $1.5 \times 10^{-22}$ & $1.6 \times 10^{45}$ & 2674 & $1.6 \times 10^{-22}$ & $1.8 \times 10^{45}$ \\
  \hline
  750 & 150 & 2674 & $2.0 \times 10^{-22}$ & $8.1 \times 10^{45}$ & 2674 & $2.2 \times 10^{-22}$ & $9.6 \times 10^{45}$ \\
  - & 400 & 2674 & $2.2 \times 10^{-22}$ & $9.9 \times 10^{45}$ & 2674 & $2.4 \times 10^{-22}$ & $1.1 \times 10^{46}$ \\
  \hline
  1550 & 150 & 2674 & $5.7 \times 10^{-22}$ & $2.7 \times 10^{47}$ & 2674 & $5.6 \times 10^{-22}$ & $2.6 \times 10^{47}$ \\
  - & 400 & 2674 & $4.9 \times 10^{-22}$ & $2.0 \times 10^{47}$ & 2674 & $5.0 \times 10^{-22}$ & $2.1 \times 10^{47}$ \\
  \hline
\end{tabular}
  \caption{\label{tab:long_upper_limits_2674} Upper limits for \ac{hrss} and \ac{EGW} estimated at 50\% detection efficiency for half sine-Gaussian and ringdown waveforms for the most sensitive burst from Swift J1818.0$-$1607 for the long-duration search.}
\end{center}
\end{table*}

\section{Conclusions}\label{sec:discussion}

In this study, we search for and find no evidence of \acp{GW} coincident with 16 bursts (13 magnetar short bursts and three electromagnetic bursts thought to be magnetar short bursts but with no identified source object) during \ac{O3}.
We search for both short-duration signals produced by excited \acp{fmode} in the neutron star's core, and for long-duration signals that may be generated from core buoyancy or Alfv\'en modes.
Statistics for both short-duration and long-duration searches are consistent with background noise; when accounting for the number of searches performed over all of the bursts, the most significant cluster found that was not clearly identified as an instrumental artifact has a p-value of 0.00857 and a $\approx29\%$ chance of being background noise.

The upper limits of both the short-duration and long-duration searches are calculated assuming optimal orientation of the source (see \appendixref{app:injections} for a discussion of the injections and upper limit calculations, and see \cite{Abbott_2019} for a discussion of the effects of the polarization on the upper limits).
For the short-duration search on bursts from known sources, the energy at $50\%$ detection efficiency \egwulF{} ranges between \unit[$ 1 \times 10 ^{46}$]{erg} and \unit[$ 2 \times 10^{47}$]{erg} for waveforms with frequencies between \unit[1500]{Hz} and \unit[2020]{Hz}. In \tableref{tab:short_duration_injections_1935_50p_ULboxes}, \tableref{tab:short_duration_injections_1818_50p_ULboxes}, and \tableref{tab:short_duration_injections_unknown_50p_ULboxes} we report the most constrained \hrssulF{} values over all bursts for the different tested signals. The only injected waveforms with exactly the same parameters as the O2 search are the white noise bursts, and we see a factor of improvement in \hrssulF{} of 1.1 for the 100$-$200 \si{\Hz}, 11 \si{\ms} WNB.
To obtain an approximate metric of improvement, we compare injected sine-Gaussian waveforms at \unit[1600]{Hz} and \unit[2020]{Hz} from \ac{O3} to sine-Gaussian waveforms of \unit[1500]{Hz} and \unit[2000]{Hz} in \ac{O2} and see factors of improvement of 1.5 and 1.7 in \hrssulF, respectively.

The long-duration search sets the lowest upper limits on long-duration \ac{GW} emission from magnetar bursts to date.
The lowest strain upper limits were for SGR 1935$+$1254, with \ac{hrss} as low as \unit[$8.7 \times 10^{-23}$]{$/\sqrt{\unit{Hz}}$} at \unit[450]{Hz} (\egwulF = \unit[$1.9 \times 10^{45}$]{erg}) for burst 2656. The lowest gravitational-wave energy upper limits come from Swift J1818.0$-$1607 because of its closer proximity (\unit[4.8]{kpc} vs \unit[9]{kpc}), with \egwulF = \unit[$2.8 \times 10^{43}$]{erg} at \unit[55]{Hz} (\hrssulF = \unit[$1.6 \times 10^{-22}$]{$/\sqrt{\unit{Hz}}$}) for burst 2674.

We also place constraints on the ratio of gravitational-wave energy to electromagnetic energy emitted by SGR 1935$+$2154 (the only source whose bursts have published electromagnetic energies) using the calculated isotropic electromagnetic energies given in \tableref{tab:table_triggers}.
The most constraining upper limits to this ratio for both the short-duration and long-duration searches come from burst 2656. 
For the short-duration search, the most constraining ratio when taking the gravitational-wave energy from the $1590 \si{\Hz}$, $100 \si{\ms}$ ringdown waveform is $\Eratioflat = 6.4 \times 10^{6}$.
For the long-duration search, the best upper limit on the ratio is
$\Eratioflat = 3.3 \times 10^3$,
which comes from a half sine-Gaussian at \unit[55]{Hz} with $\tau$ = \unit[150]{s}.
These ratios are less constraining than that of SGR 1806$-$20's 2004 giant flare, $\Eratioflat = 9 \times 10^4$ for a \unit[200]{ms} ringdown waveform at \unit[1590]{Hz} \citep{PhysRevLett.101.211102} and $ \Eratioflat \approx 5$ for a band surrounding the \unit[92.5]{Hz} \ac{QPO} in the giant flare's tail \citep{PhysRevD.76.062003,PhysRevLett.101.211102}.

With the current sensitivities of the LIGO and Virgo detectors, we can now probe well below the potential energy budgets available to generate gravitational waves from catastrophic rearrangements of the star's internal magnetic field~\citep{Ioka2001, corsi_owen_2011}.
However, even the most constraining upper limit is well above the expected \ac{GW} energy one would expect from \ac{fmode} emission from giant flares~\cite[e.g.,][]{levin07,zink12,ciolfi12}, let alone the lower-energy bursts being considered here.
As \ac{GW} observatories continue to improve in sensitivity, and more observatories such as KAGRA~\citep{KAGRA} reach comparable sensitivity, searches for \acp{GW} from magnetar bursts will eventually probe several orders of magnitude below the electromagnetic energy of giant flares, increasing the probability of a discovery of \acp{GW} from magnetar flares.

\section*{Acknowledgements}

This material is based upon work supported by NSF’s LIGO Laboratory which is a major facility
fully funded by the National Science Foundation.
The authors also gratefully acknowledge the support of
the Science and Technology Facilities Council (STFC) of the
United Kingdom, the Max-Planck-Society (MPS), and the State of
Niedersachsen/Germany for support of the construction of Advanced LIGO 
and construction and operation of the GEO\,600 detector. 
Additional support for Advanced LIGO was provided by the Australian Research Council.
The authors gratefully acknowledge the Italian Istituto Nazionale di Fisica Nucleare (INFN),  
the French Centre National de la Recherche Scientifique (CNRS) and
the Netherlands Organization for Scientific Research (NWO), 
for the construction and operation of the Virgo detector
and the creation and support  of the EGO consortium. 
The authors also gratefully acknowledge research support from these agencies as well as by 
the Council of Scientific and Industrial Research of India, 
the Department of Science and Technology, India,
the Science \& Engineering Research Board (SERB), India,
the Ministry of Human Resource Development, India,
the Spanish Agencia Estatal de Investigaci\'on (AEI),
the Spanish Ministerio de Ciencia e Innovaci\'on and Ministerio de Universidades,
the Conselleria de Fons Europeus, Universitat i Cultura and the Direcci\'o General de Pol\'{\i}tica Universitaria i Recerca del Govern de les Illes Balears,
the Conselleria d'Innovaci\'o, Universitats, Ci\`encia i Societat Digital de la Generalitat Valenciana and
the CERCA Programme Generalitat de Catalunya, Spain,
the National Science Centre of Poland and the European Union – European Regional Development Fund; Foundation for Polish Science (FNP),
the Swiss National Science Foundation (SNSF),
the Russian Foundation for Basic Research, 
the Russian Science Foundation,
the European Commission,
the European Social Funds (ESF),
the European Regional Development Funds (ERDF),
the Royal Society, 
the Scottish Funding Council, 
the Scottish Universities Physics Alliance, 
the Hungarian Scientific Research Fund (OTKA),
the French Lyon Institute of Origins (LIO),
the Belgian Fonds de la Recherche Scientifique (FRS-FNRS), 
Actions de Recherche Concertées (ARC) and
Fonds Wetenschappelijk Onderzoek – Vlaanderen (FWO), Belgium,
the Paris \^{I}le-de-France Region, 
the National Research, Development and Innovation Office Hungary (NKFIH), 
the National Research Foundation of Korea,
the Natural Science and Engineering Research Council Canada,
Canadian Foundation for Innovation (CFI),
the Brazilian Ministry of Science, Technology, and Innovations,
the International Center for Theoretical Physics South American Institute for Fundamental Research (ICTP-SAIFR), 
the Research Grants Council of Hong Kong,
the National Natural Science Foundation of China (NSFC),
the Leverhulme Trust, 
the Research Corporation, 
the Ministry of Science and Technology (MOST), Taiwan,
the United States Department of Energy,
and
the Kavli Foundation.
The authors gratefully acknowledge the support of the NSF, STFC, INFN and CNRS for provision of computational resources.

This work was supported by MEXT, JSPS Leading-edge Research Infrastructure Program, JSPS Grant-in-Aid for Specially Promoted Research 26000005, JSPS Grant-in-Aid for Scientific Research on Innovative Areas 2905: JP17H06358, JP17H06361 and JP17H06364, JSPS Core-to-Core Program A. Advanced Research Networks, JSPS Grant-in-Aid for Scientific Research (S) 17H06133 and 20H05639 , JSPS Grant-in-Aid for Transformative Research Areas (A) 20A203: JP20H05854, the joint research program of the Institute for Cosmic Ray Research, University of Tokyo, National Research Foundation (NRF), Computing Infrastructure Project of KISTI-GSDC, Korea Astronomy and Space Science Institute (KASI), and Ministry of Science and ICT (MSIT) in Korea, Academia Sinica (AS), AS Grid Center (ASGC) and the Ministry of Science and Technology (MoST) in Taiwan under grants including AS-CDA-105-M06, Advanced Technology Center (ATC) of NAOJ, and Mechanical Engineering Center of KEK.

This project has made use of the data of the Interplanetary Network (ssl.berkeley.edu/ipn3/index.html), maintained by K. Hurley. We like to remember all of the contributions K. Hurley made to many LIGO-Virgo-KAGRA searches over the years.

We would like to thank all of the essential workers who put their health at risk during the COVID-19 pandemic, without whom we would not have been able to complete this work.

\appendix

\section{Modifications to the Parameters of the Short-Duration Search}
\label{app:short-duration-parameters}

It should be noted that the burst times are distributed such that the standard \unit[3]{h} symmetric background in the short-duration searches would in some cases include the time of the previous or subsequent burst. We mitigate this by either reducing the background length or adjusting the background asymmetry factor (the fraction of background time before the burst) to optimize the amount of background data that could be used for each burst. Full details of these modifications are provided in Table \ref{tab:short_param_modifications}.

In addition to modifying the length and background asymmetry factor of the background to exclude neighboring bursts, there are also modifications to the search that we do in order to increase the sensitivity of each burst. These include vetoing events in specific frequency bands that display high non-Gaussianity and introducing an error region of the source to adjust for the motion of the Earth during the on-source window. A full list of these changes is given in Table \ref{tab:short_param_modifications}.

\renewcommand{\arraystretch}{1.5}

\begin{table*}
\begin{center}
\begin{tabular}{| c c c | }
  \hline
  Burst & Parameters of delayed on-source search & Parameters of centered on-source search \\[1ex] %
  \hline\hline
  2652 & \thead{Background Asymmetry Factor = .3726\\Background length = 8090s\\Frequency Range = 300Hz to 4000Hz} & \thead{Background Asymmetry Factor = .3858\\Background length = 9056s}  \\
  \hline
  2653 & \thead{Background Asymmetry Factor = .4711\\Background length = 10790s\\Frequency Range = 65Hz to 4000 Hz\\Error Region = 1 deg.} &    \\
  \hline
  2655 & \thead{Background Asymmetry Factor = 0.5677\\Background length = 10790s} & \thead{Background Asymmetry Factor = .5225\\Background length = 10790s}  \\
  \hline
  2656 & \thead{Background Asymmetry Factor = 0.4331\\Background length = 10790s} & \thead{Background Asymmetry Factor = .4775\\Background length = 10790s}  \\
  \hline
  2660 & \thead{Background Asymmetry Factor = 0.3421\\Background length = 10790s} & \thead{Background Asymmetry Factor = .3403\\Background length = 10790s}  \\
  \hline
  2661 & Error Region = 1 deg. &   \\
  \hline
  2668 & Error Region = 1 deg. &  Frequency Range = 85 - 4000Hz \\
  \hline
  2669 & \thead{Background Asymmetry Factor = 0.8993\\Background length = 10790s} & \thead{Background Asymmetry Factor = .854\\Background length = 10790s}  \\
  \hline
  2670 & \thead{Background Asymmetry Factor = 0.1015\\Background length = 10790s} & \thead{Background Asymmetry Factor = .1459\\Background length = 10790s}  \\
  \hline
  2671 & & \thead{Background length = 10790s\\}  \\
  \hline
  2673 & \thead{Frequency Range = 60Hz to 4000Hz\\Error Region = 1 deg.\\ } &   \\
  \hline
\end{tabular}
  \caption{\label{tab:short_param_modifications} Parameters used in the centered on-source and delayed on-source short-duration searches.  When no value is specified, the search was run with the default parameters, including frequency ranging from 50-4000 Hz, a symmetric background window 10800s in length, and 0 degree error region. The background asymmetry factor is defined as the fraction of the background time before the burst time, with 0.5 corresponding to a symmetric background. The error region is defined as the $1\sigma$ uncertainty in the sky position of the source. Using a non-zero error region on a point source can sometimes increase the search sensitivity because it counters the effects of the earth's rotation during the on-source window.
  }
\end{center}
\end{table*}

\section{Data Removed from Long-Duration Search Windows}\label{app:long-duration-data}

Data is required both before and after each pixel in the long-duration time-frequency map to estimate the pixel background (see \cite{PhysRevD.83.083004} for details). In this search, we used \unit[18]{s} of data before and after each pixel, as was done in \cite{quitzowjames2016thesis, Quitzow_James_2017, Abbott_2019}.
Thus, any data removed from the long-duration on-source window includes up to an additional \unit[36]{s}, with \unit[18]{s} both before and after the interval to be removed (we also note that \unit[1640]{s} of data are required for the full \unit[1604]{s} long-duration on-source window).

Two on-source windows were missing data.
One was the on-source window of burst 2651, which starts \unit[87]{s} after the burst; since the on-source window starts \unit[4]{s} before the burst and \unit[18]{s} is needed to estimate the pixel background, the time-frequency map starts \unit[109]{s} after the start of the on-source window (\unit[105]{s} after the burst).
Data is available for the first \unit[1121]{s} of the on-source window of burst 2665, leading to the time-frequency map ending after \unit[1103]{s}.
The on-source windows for bursts 2652, 2660 and 2665 each had \unit[8]{s} of data removed due to data quality issues, leading to gaps of \unit[44]{s} in the time-frequency maps.
As was done in previous searches (including the \ac{O2} search \citep{Abbott_2019}), noisy spectral lines, such as \si{60 \Hz} power line harmonics, were identified and removed from the time-frequency maps for each detector pair.
Of special note, 55 Hz and 150 Hz were removed for the \ac{LHO}/Virgo detector pair and 150 Hz for \ac{LLO}/Virgo; thus, these detector pairs were not sensitive to injected waveforms in these respective frequencies.
The Bezier curves for the clusters were generated identically to the other windows, with the missing times (and data removed due to noisy lines) not included in the calculation of the cluster SNR.
The background segments were treated identically to their respective on-source windows.

\section{Injected waveforms and upper limits}\label{app:injections}

The long-duration search injects all waveforms (half sine-Gaussians and ringdowns) with optimal orientation (inclination angle $\iota = 0$).
For the short-duration search, the waveforms are injected uniformly over $\cos{\iota}$ from $-1$ to $1$;
however, when calculating the upper limits \hrssulF, we treat the injections as if they all had optimal orientation of the source ($\iota = 0$), which corresponds to the lowest possible \egwulF{} from the source.

For the short-duration search upper limits, \ac{EGW} is calculated from the \ac{hrss} using the approximation for a narrow-band source with optimal orientated from equation 17 of \citep{Patricks_ROT}:
\begin{equation}
E_{GW} \approx \frac{2}{5}\frac{c^3 \pi^2}{G}d^2f_0^2 h_\mathrm{rss}^2
\;,
\end{equation}
where $d$ is the distance to the source and $f_0$ is the central frequency.

The \ac{hrss} of a sine-Gaussian is \citep{quitzowjames2016thesis}:
\begin{equation}\label{hrss-sg1}
  h_\mathrm{rss}^\mathrm{SG} =
  h_0 \tau^{1/2}
  \frac{\pi^{1/4}}{2\left(2^{1/4}\right)}
  \sqrt{
  \left(
  \frac{\left(1 + \cos^2\iota\right)^2}{4} + \cos^2{\iota}
  \right)
  +
  \left(
  \frac{\left(1 + \cos^2\iota\right)^2}{4} - \cos^2{\iota}
  \right)\euler^{-2\pi^2f_0^2\tau^2}
  }
  \;,
\end{equation}
where $\iota$ is the inclination angle and $\tau$ is the characteristic time.
The \ac{hrss} of a sine-Gaussian when the inclination angle $\iota = 0$ is:
\begin{equation}\label{hrss-sg2}
  h_{\mathrm{rss},\iota=0}^\mathrm{SG} =
  h_0
  \sqrt{\tau}
  \frac{\pi^{1/4}}{2^{3/4}}
  \;.
\end{equation}

The \ac{EGW} of a sine-Gaussian waveform is \citep{quitzowjames2016thesis}:
\begin{equation}\label{EGW-sg1}
  E_\mathrm{GW}^\mathrm{SG} =
  \frac{c^3\pi^{5/2}}{5\sqrt{2} \, G}
  h_0{}^2 d^2 f_0{}^2\tau
  \left[
    1 + \frac{1}{4\pi^2f_0^2\tau^2}
    \left(
      1 + \frac{1}{6}\euler^{-2\pi^2f_0^2\tau^2}
    \right)
  \right]
  \;.
\end{equation}
For $Q = \sqrt{2} \pi f_0 \tau \gg 1$, this can be approximated as \citep{quitzowjames2016thesis}:
\begin{equation}\label{EGW-sg2}
  E_\mathrm{GW}^\mathrm{SG} \approx
\frac{c^3\pi^{5/2}}{5\sqrt{2} \, G}
h_0{}^2 d^2 f_0{}^2\tau\;.
\end{equation}

The \ac{EGW} of a half sine-Gaussian is half of the \ac{EGW} of a sine-Gaussian, and the \ac{hrss} of a half sine-Gaussian is the \ac{hrss} of a sine-Gaussian divided by $\sqrt{2}$. 
The \ac{hrss} of a half sine-Gaussian with $\iota = 0$ is \citep{quitzowjames2016thesis}:
\begin{equation}\label{hrss-sg3}
  h_{\mathrm{rss},\iota=0}^\mathrm{hSG} = 
  \frac{h_{\mathrm{rss},\iota=0}^\mathrm{SG}}{\sqrt{2}} =
  h_0
  \sqrt{\tau}
  \frac{\pi^{1/4}}{2^{5/4}}
  \;.
\end{equation}

The \ac{EGW} of a half sine-Gaussian waveform with $Q = \sqrt{2} \pi f_0 \tau \gg 1$ can be approximated as \citep{quitzowjames2016thesis}:
\begin{equation}\label{EGW-sg3}
  E_\mathrm{GW}^\mathrm{hSG} \approx
\frac{c^3\pi^{5/2}}{10\sqrt{2} \, G}
h_0{}^2 d^2 f_0{}^2\tau\;.
\end{equation}

The \ac{hrss} of a ringdown waveform can be derived from \eqref{hrss_eq}:
\begin{equation}\label{hrss_ringdown}
  h_\mathrm{rss}^\mathrm{ringdown} =
  \frac{h_0}{2}
  \sqrt{\frac{\tau}{2}}
  \left[
  \left(\frac{(1+\cos^2{\iota})^2}{4} + \cos^2{\iota}\right)
  +
  \frac{1}{1 + 4\pi^2 f_0^2 \tau^2}\left(\frac{(1+\cos^2{\iota})^2}{4}
  - \cos^2{\iota}\right)
  \right]^{1/2}
  \;.
\end{equation}
For $\iota = 0$ this becomes:
\begin{equation}
  h_{\mathrm{rss},\iota=0}^\mathrm{ringdown} = \frac{h_0}{2}
  \sqrt{\tau}
  \;.
\end{equation}

And \ac{EGW} can be calculated to be:
\begin{equation}
  E_\mathrm{GW}^\mathrm{ringdown} =
  \frac{c^3}{40 G} h_0^2
  d^2
  \left(
  \frac{1
  +
  4\pi^2 f_0^2 \tau^2}{\tau}
  \right)
  \left(
  1
  +
  \left(\frac{1}{6}\right)
  \frac{1}{1 + 4\pi^2 f_0^2 \tau^2}
  \right)
  \;.
\end{equation}
For $Q = \sqrt{2} \pi f_0 \tau \gg 1$:
\begin{equation}
  E_\mathrm{GW}^\mathrm{ringdown} \approx
  \frac{\pi^2c^3}{10 G}
  d^2
  f_0^2
  h_0^2
  \tau
  \;.
\end{equation}

It is important to note that the short-duration ringdown waveforms include a \textit{ringup} right before the injection time for the purpose of avoiding a discontinuous jump in the signal.
This ringup has a rise time that is $\frac{1}{10}$ of the ringdown damping time.
When including this ringup, the \ac{hrss} of the total waveform is the \ac{hrss} of the ringup and ringdown added in quadrature
while
the \ac{EGW} of the ringup and ringdown are add linearly.
This gives:
\begin{equation}
  h_{\mathrm{rss},\iota=0}^\mathrm{ringup+ringdown} 
  = 
  \frac{h_0}{2}
  \sqrt{\tau + \tau \frac{1}{10}}
  = 
  \frac{h_0}{2}
  \sqrt{\tau \frac{11}{10}}
  \;,
\end{equation}
and (for $\sqrt{2} \pi f_0 \tau \gg 1$):
\begin{equation}
  E_\mathrm{GW}^\mathrm{ringup+ringdown} \approx
  \frac{\pi^2c^3}{10 G}
  d^2
  f_0^2
  h_0^2
  \tau (11/10)
  \;.
\end{equation}

\bibliography{bibliography}

\let\author\myauthor
\let\affiliation\myaffiliation
\let\maketitle\mymaketitle
\newpage
\section*{Authors}
\title{Authors}


\author{R.~Abbott}
\affiliation{LIGO Laboratory, California Institute of Technology, Pasadena, CA 91125, USA}
\author{H.~Abe}
\affiliation{Graduate School of Science, Tokyo Institute of Technology, Meguro-ku, Tokyo 152-8551, Japan  }
\author{F.~Acernese}
\affiliation{Dipartimento di Farmacia, Universit\`a di Salerno, I-84084 Fisciano, Salerno, Italy  }
\affiliation{INFN, Sezione di Napoli, Complesso Universitario di Monte S. Angelo, I-80126 Napoli, Italy  }
\author[0000-0002-8648-0767]{K.~Ackley}
\affiliation{OzGrav, School of Physics \& Astronomy, Monash University, Clayton 3800, Victoria, Australia}
\author[0000-0002-4559-8427]{N.~Adhikari}
\affiliation{University of Wisconsin-Milwaukee, Milwaukee, WI 53201, USA}
\author[0000-0002-5731-5076]{R.~X.~Adhikari}
\affiliation{LIGO Laboratory, California Institute of Technology, Pasadena, CA 91125, USA}
\author{V.~K.~Adkins}
\affiliation{Louisiana State University, Baton Rouge, LA 70803, USA}
\author{V.~B.~Adya}
\affiliation{OzGrav, Australian National University, Canberra, Australian Capital Territory 0200, Australia}
\author{C.~Affeldt}
\affiliation{Max Planck Institute for Gravitational Physics (Albert Einstein Institute), D-30167 Hannover, Germany}
\affiliation{Leibniz Universit\"at Hannover, D-30167 Hannover, Germany}
\author{D.~Agarwal}
\affiliation{Inter-University Centre for Astronomy and Astrophysics, Pune 411007, India}
% ...

\collaboration{3000}{The LIGO Scientific Collaboration, the Virgo Collaboration, and the KAGRA Collaboration}



\maketitle

\end{document}